\documentclass[twocolumn,usenames,dvipsnames]{aastex62}
\usepackage{multirow}
\usepackage{amsmath}
\usepackage{xcolor}

\newcommand{\picfull}[2]{
	\begin{figure*}
     \plotone{#1}
	 \caption{#2\label{#1}}
	\end{figure*}}
\newcommand{\picfullwide}[2]{
	\begin{figure*}
     \includegraphics[width=\textwidth]{#1}
	 \caption{#2
	 \label{#1}}
	\end{figure*}}
\newcommand{\picwide}[2]{
	\begin{figure}
	 \includegraphics[width=\columnwidth]{#1}
	 \caption{#2}
	 \label{#1}
	\end{figure}}
\graphicspath{{./}{figures/}}

\accepted{May 4, 2020}
\submitjournal{ApJ}

\shorttitle{Modeling K2-18b with 1D-TERRA}
\shortauthors{Scheucher et al.}

\begin{document}

\correspondingauthor{Markus Scheucher}
\email{scheucher@tu-berlin.de, markus.scheucher@dlr.de}

\author[0000-0003-4331-2277]{Markus Scheucher}\thanks{Equal Contribution Authors}
\affiliation{Zentrum f\"{u}r Astronomie und Astrophysik, Technische Universit\"{a}t Berlin, 10623 Berlin, Germany}
\affiliation{Institut f\"{u}r Planetenforschung, Deutsches Zentrum f\"{u}r Luft- und Raumfahrt, 12489 Berlin, Germany}

\author[0000-0002-2238-5269]{F.~Wunderlich}\thanks{Equal Contribution Authors}
\affiliation{Zentrum f\"{u}r Astronomie und Astrophysik, Technische Universit\"{a}t Berlin, 10623 Berlin, Germany}
\affiliation{Institut f\"{u}r Planetenforschung, Deutsches Zentrum f\"{u}r Luft- und Raumfahrt, 12489 Berlin, Germany}

\author{J.L.~Grenfell}
\affiliation{Institut f\"{u}r Planetenforschung, Deutsches Zentrum f\"{u}r Luft- und Raumfahrt, 12489 Berlin, Germany}

\author{M.~Godolt}
\affiliation{Zentrum f\"{u}r Astronomie und Astrophysik, Technische Universit\"{a}t Berlin, 10623 Berlin, Germany}

\author[0000-0001-7196-6599]{F.~Schreier}
\affiliation{Institut f\"{u}r Methodik der Fernerkundung, Deutsches Zentrum f\"{u}r Luft- und Raumfahrt,  82234 Oberpfaffenhofen, Germany}

\author{D.~Kappel}
\affiliation{Institut f\"{u}r Planetenforschung, Deutsches Zentrum f\"{u}r Luft- und Raumfahrt, 12489 Berlin, Germany}
\affiliation{Institut f\"{u}r Physik und Astronomie, Universit\"{a}t Potsdam, 14476 Potsdam, Germany}

\author{R.~Haus}
\affiliation{Institut f\"{u}r Geowissenschaften, Universit\"{a}t Potsdam, 14476 Potsdam, Germany}

\author[0000-0001-5622-4829]{K.~Herbst}
\affil{Institut f\"ur Experimentelle und Angewandte Physik, Christian-Albrechts-Universit\"at zu Kiel, 24118 Kiel, Germany}

\author{H.~Rauer}
\affiliation{Zentrum f\"{u}r Astronomie und Astrophysik, Technische Universit\"{a}t Berlin, 10623 Berlin, Germany}
\affiliation{Institut f\"{u}r Planetenforschung, Deutsches Zentrum f\"{u}r Luft- und Raumfahrt, 12489 Berlin, Germany}
\affiliation{Institut f\"{u}r Geologische Wissenschaften, Freie Universit\"{a}t Berlin, 12249 Berlin, Germany}



\title{Consistently Simulating a Wide Range of Atmospheric Scenarios for K2-18b with a Flexible Radiative Transfer Module}

\begin{abstract}
The atmospheres of small, potentially rocky exoplanets are expected to cover a diverse range in composition and mass. Studying such objects therefore requires flexible and wide-ranging modeling capabilities. 
We present in this work the essential development steps that lead to our flexible radiative transfer module, REDFOX, and validate REDFOX for the Solar system planets Earth, Venus and Mars, as well as for steam atmospheres. REDFOX is a k-distribution model using the correlated-k approach with random overlap method for the calculation of opacities used in the $\delta$-two-stream approximation for radiative transfer. Opacity contributions from Rayleigh scattering, UV / visible cross sections and continua can be added selectively.
With the improved capabilities of our new model, we calculate various atmospheric scenarios for K2-18b, a super-Earth / sub-Neptune with $\sim$8~M$_\oplus$ orbiting in the temperate zone around an M-star, with recently observed H$_2$O spectral features in the infrared. We model Earth-like, Venus-like, as well as H$_2$-He primary atmospheres of different Solar metallicity and show resulting climates and spectral characteristics, compared to observed data. Our results suggest that K2-18b has an H$_2$-He atmosphere with limited amounts of H$_2$O and CH$_4$. Results do not support the possibility of K2-18b having a water reservoir directly exposed to the atmosphere, which would reduce atmospheric scale heights, hence too the amplitudes of spectral features inconsistent with the observations. We also performed tests for H$_2$-He atmospheres up to 50 times Solar metallicity, all compatible with the observations. 
\end{abstract}


%
\section{Introduction} \label{sec:intro}
\picfullwide{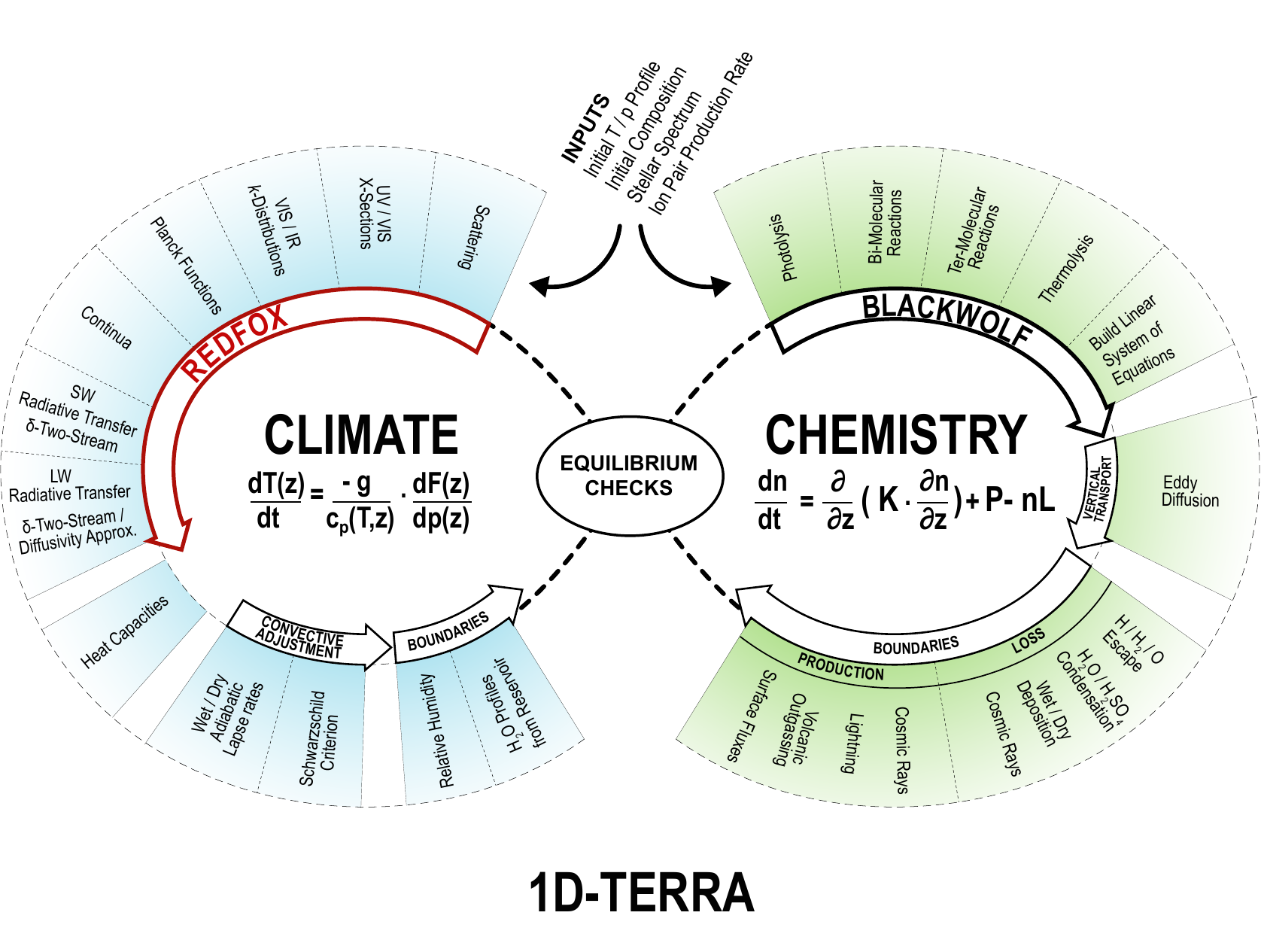}{Schematic overview of our 1D coupled climate disequilibrium-chemistry model 1D-TERRA including the new radiative transfer module, REDFOX, and the updated chemistry scheme, BLACKWOLF.}
Exciting recent discoveries in exoplanetary science include the TRAPPIST-1 system \citep{gillon2017}, Proxima Centauri b \citep{anglada2016}, and potential super-Earths / warm sub-Neptunes like LHS1140b and c \citep{ment2019}, K2-18b \citep{montet2015}, and GJ1214b \citep{charbonneau2009}. Exoplanet science is transitioning from detection into first atmospheric characterizations from spectral observations, giving us insights into their composition and possible formation and evolution. Numerous spectral observations of Jupiter-sized to warm Neptune-sized planets have been reported and discussed in recent years \citep[see, e.g.,][]{sing2016,crossfield2017}, but the first detection of water features in the atmosphere of the temperate (T$\rm _{eq}\sim 272~K$) super-Earth / sub-Neptune K2-18b \citep{benneke2019,Tsiaras2019} orbiting an early-type M-star is especially exciting, as it offers an unprecedented possibility to gain insights into the atmosphere and climate of objects in the regime between rocky and gas planets which do not exist in the Solar System. Detailed understanding of atmospheric processes, such as radiative transfer, convection and disequilibrium chemistry is key for the interpretation of such spectral detections. \\
The atmospheres of terrestrial planets lying in the habitable zone of their respective host stars could be from H$_2$-He dominated, to H$_2$O, CO$_2$ and N$_2$ dominated, or even O$_2$ or CO dominated for warmer and cooler planets respectively \citep[see, e.g.,][]{madhusudhan2016,forget2014}. \\

A main motivation for our model development is as follows. The radiative transfer schemes, based on the k-distribution method, implemented in previous versions of our model as well as other similar models developed for the study of terrestrial planets \citep[e.g.][]{kopparapu2013,segura2010}, rely on pre-mixed k-tables for atmospheric conditions, such as pressure, temperature, and composition. Especially the latter can be a considerable restriction for simulating atmospheres that are more and more different from that of Earth. Adding radiation-absorbing constituents often requires a recalculation of all k-tables. The same applies for including updates of line lists, for example, for one constituent. There are, however, multiple ways of treating overlap of spectral absorption lines of gaseous components in k-distribution radiative transfer calculations for more flexibility, assuming perfect correlation, random overlap, or disjoint lines \citep[see, e.g.,][]{pierrehumbert2010}. \citet{lacis1991} have described how the overlap of absorption by gaseous components can be treated quickly and accurately, using the random overlap assumption in the k-distribution method, which has been implemented and tested for hot Jupiter studies by \citet{amundsen2016}. \citet{Malik_2017} have recently developed a GPU-based open source radiative transfer model using the faster assumption of perfect correlation between spectral lines of different molecules in the correlated-k approximation for studying hot Jupiters and other planets with primary atmospheres. This assumption, however, becomes less accurate the more absorbers are present. While k-distribution models operate in cross section-space, \citet{kitzmann2017} used opacity sampling in his CO$_2$ clouds studies, which can be seen as a degraded Line-By-Line (LBL) radiative transfer model. Then, the addition of absorbing gases is fully additive, but opacity sampling is generally computationally more expensive than the k-distribution method and becomes less accurate for lower pressures, where many thin absorption lines can be missed by wavelength discretization. \citet{Lincowski_2018} adopted a different approach by introducing a LBL radiative transfer module into a climate-chemistry model using the Linearized Flux Evolution approach \citep{ROBINSON2018} in order to reduce the number of time-consuming radiative transfer calculations for the study of terrestrial climates. This approach is fast as long as changes in important parameters influencing radiative fluxes, such as temperature and composition of major absorbers are small. In that case, fluxes are approximated by linear flux gradients stored in a Jacobian for the last calculated state, rather than invoking full radiatiative transfer calculations.\\
Our new radiative transfer model, REDFOX, using the random overlap assumption, aims to combine the flexibility with regard to stellar spectra and atmospheric conditions similar to a LBL model, but with calculation times approaching those of other k-distribution models using the correlated-k approach. A central aim of this work is to show our new extensive capabilities for studying terrestrial exoplanets with REDFOX as part of our 1D coupled climate-chemistry model, 1D-TERRA. \\
In Section \ref{sec:model}, we provide information on general climate and chemistry model updates that lead to our new 1D climate-chemistry model 1D-TERRA. Section \ref{ssec:rad} focuses in detail on the essentials of REDFOX, including the k-distribution method using the random overlap approximation. Section \ref{sec:valid} validates REDFOX against known properties of atmospheres in the Solar System. Section \ref{sec:results} presents our results for the super-Earth / sub-Neptune K2-18b before our final remarks in Section \ref{sec:sum}.
%
\section{Methodology} \label{sec:model}
Our 1D coupled climate-chemistry model has a long heritage dating back to e.g. \citet{Kasting1986, pavlov2000, segura2003} for the study of Earth-like planets. Since then, it has been extensively updated in our group, for example, \citet{rauer2011, grenfell2012, Scheucher_2018}. \citet{vonParis2008,vonParis2010,vonparis2015} implemented MRAC, a modified version of the Rapid Radiative Transfer Model (RRTM) \citep{mlawer1997} for CO$_2$ dominated atmospheres in the climate module to study e.g. early Mars or planets at the outer edge of the habitable zone. Our new radiative transfer module, REDFOX, is designed to operate over a wide range of stellar energy spectra, as well as diverse neutral composition  (without ion chemistry) and pressure-temperature conditions in terrestrial atmospheres. \\
\citet{grenfell2007}, \citet{tabataba2016}, and \citet{Scheucher_2018} implemented parameterizations of cosmic rays and stellar energetic particles into our chemical solver. The latter can be part of an extensive model suite, calculating the precipitation of energetic particles through a magnetosphere and atmosphere, induced atmospheric ionization and the impact on climate and neutral atmospheric composition, as described in \citet{herbst2019}. In conjunction with the companion paper \citet[][in prep.]{wunderlich2019b}, we briefly describe in Subsection \ref{ssec:chem} key aspects of our extensively updated chemical reaction scheme which can simulate diverse terrestrial atmospheres. \\
The new model with updated climate and chemistry will be referred to as 1D-TERRA. Fig. \ref{1D-TERRA} provides a schematic overview of the 1D-TERRA model.\\
For post-processing we use the "Generic  Atmospheric  Radiation  Line-by-line  Infrared  Code" GARLIC \citep[e.g.][]{schreier2014, schreier2018agk, schreier2018ace} to calculate planetary transit and emission spectra as described in e.g. \citet{Scheucher_2018, wunderlich2019a}. 
%
\subsection{Radiative Transfer Module REDFOX} \label{ssec:rad}
\picfullwide{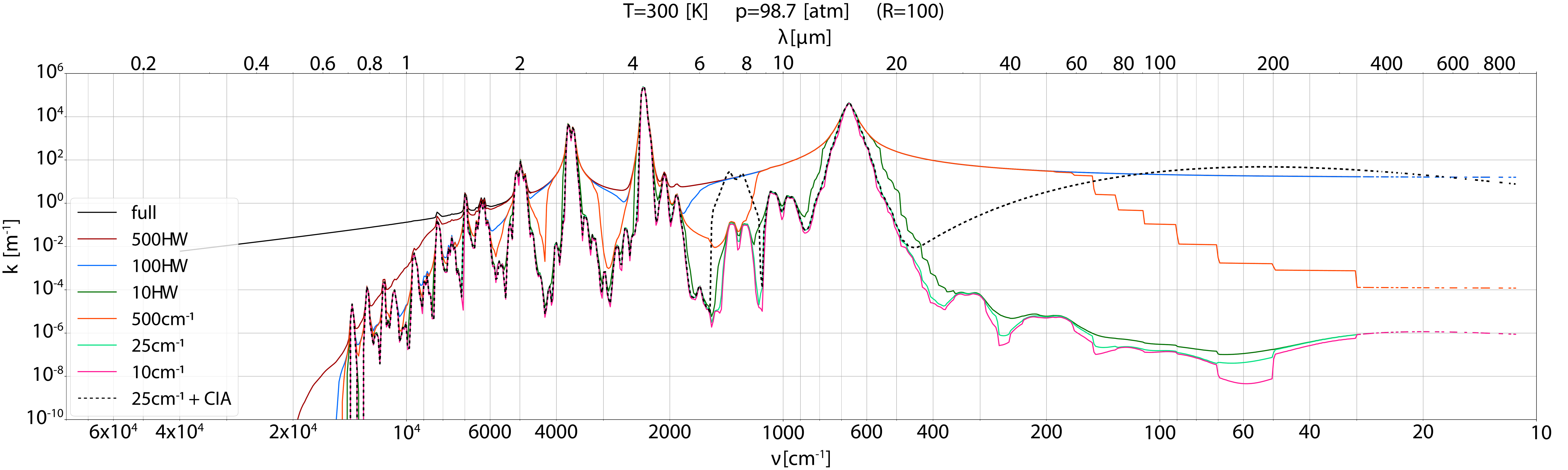}{Comparison of CO$_2$ absorption coefficients calculated with KSPECTRUM from HITRAN2016 with sub-Lorentzian wings after \citet{PERRIN1989311} at T~=~300~K and p~=~100~bar for different line-wing cut-off values (solid lines). We compare cut-off values in HW (= line Half-Widths, $\gamma_\nu$) and in fixed wavenumber values (cm$^{-1}$). Overplotted (black dashed) are results for a line-wing cut-off at 25~cm$^{-1}$ including CIA parameterizations from \citet{GRUSZKA1997172} and \citet{BARANOV2004432}.}
REDFOX calculates radiative transfer for discrete intervals, or bands, in the spectral range from $\nu=0\rm\,cm^{-1}$ to $\nu=10^5\rm\,cm^{-1}$, where any of the spectral intervals can be used for the short-wave (SW) and long-wave (LW) treatment of radiative transfer to allow e.g. for a more complete coverage of the irradiation of late M-dwarfs. Also, in the LW the choice of the spectral interval is now flexible to include colder, temperate, and warmer planets alike. In addition we extended the pressure and temperature range for the correlated-k calculations so that absorption by, for example, O$_3$, H$_2$O, CO$_2$, and CH$_4$ can be treated also for planets close to the inner edge of the habitable zone, where evaporation of water can lead to surface pressures higher than 1.5 bar, which was our previous pressure limit for Earth-like atmospheres.  Note that \citet{vonParis2010} implemented absorption by CO$_2$ and H$_2$O for pressures up to 1000~bar for a high-CO$_2$ atmospheric study without trace-gases in an earlier model version. In contrast, our extended range now includes absorption by all molecules presented in Table~\ref{tbl:molecules}, which e.g. also allows for modeling  steam atmospheres, habitable-zones, and the upper atmospheres of sub-Neptunes, or other H$_2$-dominated atmospheres. 
%
\subsubsection{Calculation of VIS / IR Cross Sections}
\begin{deluxetable*}{l l}
    \tabletypesize{\footnotesize}
    \tablecolumns{2}
    \tablecaption{List of molecules included in REDFOX and the corresponding VIS / IR line list and  UV / VIS cross section sources. \label{tbl:molecules}}
    \tablehead{\colhead{Data Source} & \colhead{Molecules}}    
    \startdata
         \multicolumn{1}{p{0.25\textwidth}}{\multirow{2}{*}{HITRAN 2016 line list\tablenotemark{a}}} & \multicolumn{1}{p{0.65\textwidth}}{CH$_3$Cl, CH$_4$, CO, CO$_2$, H$_2$, H$_2$O, HCl, HCN, HNO$_3$, HO$_2$, HOCl, N$_2$, N$_2$O, NH$_3$, NO, NO$_2$, O$_2$, O$_3$, OH, SO$_2$} \\ \hline
         \multicolumn{1}{p{0.25\textwidth}}{\multirow{5}{*}{MPI Mainz Spectral Atlas\tablenotemark{b}}} & \multicolumn{1}{p{0.65\textwidth}}{C$_2$H$_2$, C$_2$H$_2$O, C$_2$H$_3$, C$_2$H$_4$, C$_2$H$_4$NH, C$_2$H$_5$, C$_2$H$_5$CHO, C$_2$H$_6$, C$_3$H$_3$, C$_3$H$_6$, C$_3$H$_8$, C$_4$H$_2$, CH$_2$CCH$_2$, CH$_2$CO, CH$_3$,  CH$_3$C$_2$H, CH$_3$CHO, CH$_3$Cl, CH$_3$NH$_2$, CH$_3$OH, CH$_3$ONO, CH$_3$ONO$_2$, CH$_3$OOH, CH$_4$, Cl$_2$, Cl$_2$O, Cl$_2$O$_2$, ClCO$_3$, ClO, ClONO, ClONO$_2$, ClOO, ClS$_2$, CO, CO$_2$, COCl$_2$, CS$_2$, H$_2$, H$_2$CO, H$_2$O, H$_2$O$_2$, H$_2$S, H$_2$SO$_4$, HCl, HCN, HCO, HCOOH, HNO, HNO$_2$, HNO$_3$, HO$_2$, HO$_2$NO$_2$, HOCl, HSO, N$_2$, N$_2$H$_2$, N$_2$H$_4$, N$_2$O, N$_2$O$_5$, NH$_3$, NO, NO$_2$, NO$_3$, NOCl, O$_2$, O$_3$, OClO, OCS, OH, S$_2$, S$_2$O, S$_2$O$_2$, S$_3$, S$_4$, SCl, SCl$_2$,  SNO, SO, SO$_2$, SO$_2$Cl$_2$, SO$_3$} \\
    \enddata
    \tablenotetext{a}{\url{www.hitran.org/LBL/} \citep{GORDON20173}}
    \tablenotetext{b}{\url{http://satellite.mpic.de/spectral_atlas/cross_sections/} \citep{spectralatlas}}
\end{deluxetable*}
In the random overlap method, cross sections calculated for each gas constituent separately are then cross-correlated in the radiative transfer calculations of atmospheric transmission functions. For the calculation of spectroscopic cross sections we use the HITRAN 2016 line list \citep{GORDON20173} and the open-source KSPECTRUM code originally developed by \citet{Eymet_2016}. We  downloaded KSPECTRUM1.2.0\footnote{downloaded from \url{https://www.meso-star.com/projects/art/kspectrum.html}} in 2017, included the treatment of HITRAN 2016 in KSPECTRUM and also implemented the "Total Internal Partition Sums" (TIPS) from \citet{GAMACHE201770}, used to calculate temperature conversion factors applied to line intensities, into what we now refer to as  KSPECTRUM\_Htr16. In 2018, KSPECTRUM1.3 was released with updates on CO$_2$ treatment. These updates, e.g., removed the possibility of line truncation when sub-Lorentzian wings are used. Since we decided to cut line-wings and add missing contributions from far wings with collision-induced absorption, these modifications are not included in our calculations. A short discussion on our approach follows below. Our KSPECTRUM\_Htr16 source code is available on Github\footnote{KSPECTRUM\_Htr16: \url{https://<placeholder>} \textbf{to be updated before publication}}.\\
For the spectral discretization of the Voigt profiles we define line center regions up to 10 half-widths (HW) from the nominal line center wavenumbers, which are calculated at 8 points per HW, calculate 16 points in the far wings, and use a cut-off at 25 cm$^{-1}$ from the line center for every molecule (see below). After extensive testing on numerous computing platforms (not shown), the above mentioned parameter set for spectral discretization represented a reasonable trade-off between accuracy and computing time ($<$  several months for low pressures). KSPECTRUM also has one discretization algorithm integrated that aims to keep the error from discretizing the Voigt profiles below a user defined value. Unfortunately this algorithm becomes computationally inefficient for low pressures where lines are very narrow, leading to a sharp increase in points needed for the wavenumber grid discretization. Our choice of discretization showed less than 10$^{-5}$  relative deviation in resulting cross sections from results using that error-limited algorithm for tests at 1 Pa (and even less deviation at higher pressures), except for H$_2$O which showed higher deviations of up to 1 percent.  It is noteworthy that calculation times for cross sections should go down significantly with the use of GPU-based algorithms, such as, e.g., Helios-k \citep{grimm2015}.
The trade-off between wavenumber discretization and accuracy is a common challenge in molecular absorption modeling, and different LBL codes use different approaches for the grid discretization and interpolation to speed up calculations, see e.g. \citet{Schreier2006} for a discussion. \\
Further, one has to be careful with regard to which line cut-off value to choose (Fig.~\ref{cut-off}).  Appropriate line-wing cut-off parameterizations together with the right pressure-broadened wing profiles are still debated in the literature \citep[see, e.g., ][]{sharp2007,Takagi2010,grimm2015,hedges2016}. It is beyond the scope of this paper to outline all proposed theories, discussions and possible assumptions surrounding this topic, instead we would like to quickly motivate our choices. Assuming we could produce all true line-wing shapes, their deviation from Lorentzian profiles and their changes with temperature, pressure and composition, adding up the contribution of all absorption lines to infinity may, theoretically, reproduce most known spectral features including absorption bands, window regions and possibly some continua. Unfortunately, many line-wings differ substantally from the classical Lorentzian approximation. \citet{clough1992} had already argued, that the super-Lorentzian nature of H$_2$O absorption is too complex to parameterize with wing shapes, and rather developed continuum parameterizations to be used in addition to line-by-line calculations with a 25~cm$^{-1}$ cut-off. The sub-Lorentzian nature of CO$_2$ line-wings has proven to be even more difficult, since numerous major absorption bands need their own parameterization \citep[see, e.g.,][]{PERRIN1989311,pollack1993,tonkov1996}. Therefore a multitude of continuum and collision-induced-absorption (CIA) parameterizations have emerged \citep[see, e.g.,][]{GRUSZKA1997172,BARANOV2004432}. \citet{sharp2007} recommended a general pressure-dependent cut-off, $d=\rm{min}(25p,100)~\rm{cm}^{-1}$, with $p$ being the atmospheric pressure in atm, leading to cut-off values $>$~25~cm$^{-1}$ for pressures $>$~1~atm and therefore being important for high-pressure atmospheric studies, such as gas planet atmospheres. \citet{wordsworth2010} conducted a thorough comparison of a 25~cm$^{-1}$ cut-off versus untruncated line-wings for CO$_2$, together with various CIA parameterizations. Following these and similar studies, some authors studying terrestrial atmospheres started adopting larger cut-off values of 500~cm$^{-1}$ \citep[e.g.][]{kopparapu2013} or even untruncated line shapes for selected molecules \citep[e.g.][]{wordsworth2010gliese,wordsworth2013}. \\
Figure~\ref{cut-off} suggests that taking into account the far wings of all absorption lines of a specific molecule over the whole wavenumber range may reproduce certain parts of continuum absorption in specific wavelength ranges, but at the same time may overestimate absorption in other spectral regions where the addition of many Lorentzian line wings from the whole absorption spectrum would create additional artificial continua not observed in experiments. For pressures above a few bar this effect can effectively mask any CO$_2$ or H$_2$O absorption features in the visible and near infrared range.  While we tested for a wide range of  p-T conditions (for CO$_2$ and H$_2$O) and various sub-Lorentzian profiles \citep{PERRIN1989311,pollack1993,tonkov1996}, in Figure \ref{cut-off} we   show this effect for CO$_2$ with sub-Lorentzian wings after \citet{PERRIN1989311} at 100 bar (98.7 atm) and 300 K for different line wing cut-off values (colors), compared to results without any cut-off (black). We also show (dotted) absorption spectra based on our 25 cm$^{-1}$ cut-offs including CO$_2$ CIAs, namely the induced dipoles after \citet{GRUSZKA1997172} and the CO$_2$-CO$_2$ dimers from \citet{BARANOV2004432}. Our results suggest that too large line cut-off values in the LBL calculations can lead to absorption coefficients far exceeding CIA continua contributions in certain spectral window regions, e.g. $8-14~\rm{\mu m}$ or $20-50~\rm{\mu m}$.  Probably the main difference to the work by \citet{wordsworth2010} is the extensive update for CO$_2$ in HITRAN2016, especially in the mid-infrared. When no cut-off is used at all (black-solid), any CO$_2$ features in the visible and near-IR are masked by the far wing contributions. This is mainly due to the previously discussed poorly understood behaviour of the far line wings. See, e.g., \citet{Takagi2010}, for a detailed discussion on the influence of pressure-broadened CO$_2$ line shapes upon radiative-convective equilibrium temperatures. \\
After testing for a range of molecules, pressures, and temperatures, we chose a line wing cut-off of 25 cm$^{-1}$ for all our calculations of k-distributions for all molecules. For CO$_2$ we calculated sub-Lorentzian wings from \citet{PERRIN1989311} within its validity range and with the same 25 cm$^{-1}$ cut-off. CIAs are added and interpolated directly in REDFOX when performing radiative transfer calculations together with other continua as described in Subsection \ref{sssec:conti}.  We would like to emphasize that we do not claim that this approach is superior to others in accuracy, but allows for flexible treatment of continua, and easy substitution when updated parameterizations or measurements become available.\\
Since we calculate absorption cross sections for pure gases of a given molecule (Table \ref{tbl:molecules}) separately,  we can only account for self-broadening and not for foreign-broadening of absorption lines.  Since pressure-broadening of any given molecule by the total atmospheric gas pressure is therefore approximated by self-broadening for that pressure environment, this could become relevant for trace species in high pressure environments for which self-broadening coefficients could differ significantly from their foreign-broadening coefficients. That said, one should note that foreign-broadening listed in HITRAN refers to (Earth) air broadening only , which should be used with caution for non Earth-like atmospheres as discussed, e.g., by \citet{wordsworth2013}. For selected molecules, there exist, however, more general pressure broadening descriptions in the Exomol database \citep{exomol}.  To tackle this problem consistently, for any species of interest would require a tabulated parameter for broadening by any possible major constituent in an atmosphere of interest. Then the line wings would have to be calculated in line-by-line calculations using e.g. the broadened half width  $\gamma = vmr_{\rm{self}}*\gamma_{\rm{self}} + \sum_i(\rm{vmr}_i*\gamma_i)$, with $vmr$ being the volume mixing ratio and $i$ any atmospheric constituent other than the considered species itself. This then needs to be tabulated in pre-mixed k-tables in $n$-dimensions for $n$ major constituents of interest, calculated on a discrete vmr grid, which would somewhat contradict our main aim here, namely a flexible radiative transfer model that is also straightforward to update with additional molecules of interest.\\
High resolution cross section lists have been calculated in this work for each of the 20 HITRAN 2016 molecules shown in Table \ref{tbl:molecules} on a pressure-temperature grid for a total of 121 points. We considered thereby 11 pressures equally spaced in log(p) ranging from 10$^{-7}$ - 10$^3$ bar or 0.01 - 10$^8$ Pa, and the following 11 temperatures: T(K) = [100, 150, 200, 250, 300, 350, 400, 500, 600, 800, 1000].  Studies calculating the outgoing longwave radiation (OLR) of hot atmospheres ($\gtrsim$ 800~K) e.g. \citet[][]{kopparapu2013,Katyal2019} have shown that these cases may require the calculation of cross sections from databases  which focus on high temperatures, such as HITEMP \citep{hitemp}  or Exomol \citep{exomol}, although our Venus validation (Section \ref{sec:valid}) suggests that this effect may not be central for the atmospheres studied here.
%
\subsubsection{k-distributions and spectral bands} \label{sssec:rorr}
\begin{deluxetable}{c >{\centering\arraybackslash}p{3.7cm} >{\centering\arraybackslash}p{3.7cm}}
    \tabletypesize{\footnotesize}
    \tablecolumns{3}
    \tablecaption{Comparison of quadrature weights  and respective g sub-intervals in brackets from RRTM ($w_{RRTM}$) and Gauss-Legendre polynomials ($w_{GL}$).\label{tbl:quads}}
    \tablehead{\colhead{i} & \colhead{$w_{RRTM}$   [$g$-bin]}  & \colhead{$w_{GL}$   [$g$-bin]}}    
    \startdata
         1 & 0.15275\quad[0.00000 - 0.15275] & 0.01358\quad[0.00000 - 0.01358] \\
         2 & 0.14917\quad[0.15275 - 0.30192] & 0.03113\quad[0.01358 - 0.04471] \\
         3 & 0.14210\quad[0.30192 - 0.44402] & 0.04758\quad[0.04471 - 0.09229] \\
         4 & 0.13169\quad[0.44402 - 0.57571] & 0.06231\quad[0.09229 - 0.15460] \\
         5 & 0.11819\quad[0.57571 - 0.69390] & 0.07480\quad[0.15460 - 0.22940] \\
         6 & 0.10193\quad[0.69390 - 0.79583] & 0.08458\quad[0.22940 - 0.31398] \\
         7 & 0.08328\quad[0.79583 - 0.87911] & 0.09130\quad[0.31398 - 0.40528] \\
         8 & 0.06267\quad[0.87911 - 0.94178] & 0.09472\quad[0.40528 - 0.50000] \\
         9 & 0.04249\quad[0.94178 - 0.98427] & 0.09472\quad[0.50000 - 0.59472] \\
         10 & 0.00463\quad[0.98427 - 0.98890] & 0.09130\quad[0.59472 - 0.68602] \\
         11 & 0.00383\quad[0.98890 - 0.99273] & 0.08458\quad[0.68602 - 0.77060] \\
         12 & 0.00303\quad[0.99273 - 0.99576] & 0.07480\quad[0.77060 - 0.84540] \\
         13 & 0.00222\quad[0.99576 - 0.99798] & 0.06231\quad[0.84540 - 0.90771] \\
         14 & 0.00141\quad[0.99798 - 0.99939] & 0.04758\quad[0.90771 - 0.95529] \\
         15 & 0.00054\quad[0.99939 - 0.99993] & 0.03113\quad[0.95529 - 0.98642] \\
         16 & 0.00007\quad[0.99993 - 1.00000] & 0.01358\quad[0.98642 - 1.00000] \\
    \enddata
\end{deluxetable}
k-distributions are cross section ($\sigma$) probability distributions within a frequency band, represented by a small number of k-coefficients and respective weighting quadratures. For a detailed description of the k-distribution method and the implementation of the Random Overlap approximation with the "Ranking and Reblocking" or "Resorting and Rebinning" RORR method, we refer to, for example, \citet{lacis1991} or \citet{amundsen2016}. The assumption behind random overlap is that absorption lines of individual molecules are uncorrelated. Although computationally more expensive than the simple addition of all molecular cross sections with their respective concentrations, this assumption is certainly a good approximation when including multiple molecules of different shapes and sizes as we do here, leading to very different roto-vibrational and translational absorption features. In random overlap, k-distributions, or $\sigma$-distributions to be more precise, need to be calculated for every pressure $p$, temperature $T$, and molecule separately. 
To calculate individual k-distributions from cross sections, we make use of the KDISTRIBUTION package, developed by V. Eymet, for easy handling of high-resolution cross section files previously generated by KSPECTRUM. While we require REDFOX to be as flexible as possible in handling different gas mixtures and atmospheric conditions, note that we will lose any frequency dependence within spectral bands when converting cross sections into k-distributions. K-coefficients of different gas mixtures within one band will not generally map to the same frequencies as in a different gas mixture. This means that we cannot directly map a given source function such as a stellar spectrum or atmospheric Planck function, or other opacity sources like Rayleigh scattering coefficients, continua, or aerosols, into individual k-coefficients. These have to be added as  band-integrated averages. For this reason, individual bands should be as small in frequency range as possible, to minimize inaccuracies introduced by using the above-mentioned band-mean values, while at the same time being fast enough for climate-evolution studies. We tested different band widths for an Earth-like atmosphere ranging from the original bands (38 for SW, 25 for LW) up to 480 bands over the whole spectral range (not shown). We chose 128 bands in total between $\nu$=0 cm$^{-1}$($\lambda$=$\infty$) and $\nu$=10$^5$ cm$^{-1}$($\lambda$=100 nm) as further increases in band numbers resulted in OLR changes of $<1\%$. Bands are evenly spaced in log($\nu$) above $\nu$=100 cm$^{-1}$, plus 10 bands with $\Delta\nu$=10 cm$^{-1}$ between $\nu$=0-100 cm$^{-1}$. The design is such that we can subsequently change the spectral bands for a specific study if accuracy or computational efficiency become an issue. Such a change requires only recalculation of k-distributions from the high resolution (8 pts/HW) cross section files, which takes approximately  a total of 24 hours on 100 CPUs on our university server for all 20 molecules  and 121 p-T points currently included. Details on band-averaged sources and opacities follow later in this section.\\
Additionally, one must choose a suitable number of quadrature weights or k-coefficients within a band and the quadrature rule. The quadrature rule used in previous versions of our code dates back to RRTM \citep{mlawer1997} and is a modification of the half Gauss-Legendre weighting, placing more emphasis on the absorption line-centers ($g \sim 1$), as described in their paper. We use those quadratures for comparison since their weighting was specifically designed to resolve well the line-center absorption for cooling rates in Earth's atmosphere. For exoplanet applications, however, we cannot generally assume that the main contribution to cooling rates arises mainly from the line-centers alone; there can be additional cooling e.g. from extensive broadening at higher pressures or temperatures. To test this, we also calculate k-distributions by applying the standard Gauss-Legendre quadrature rule using 16$\rm^{th}$ order Legendre polynomial $P_{16}$($\phi$) shifted to the [0,1] interval. For transmission $\mathcal{T}$ through a homogeneous layer this is:
\begin{equation}
    \mathcal{T} = \int_0^1 f(g)\; dg = \frac{1}{2}\sum_{i=1}^{16} w_i\; f \left(\frac{1+g_i}{2} \right) ,
\end{equation}
with $g_i$ being the i$\rm^{th}$ root of $P_{16}$  and $f=\rm{exp}(-\Delta\tau/\mu)$ where $\mu=\cos\Theta$ and $\Theta$ is the mean zenith angle used in the two-stream approximation. As given by \citet{abramowitz1972} the associated quadrature weights $w_i$ are a function of the derivatives of P$_{16}$, specifically
\begin{equation}\label{eq:lq}
    w_i = \frac{2}{(1-g_i)\;P^{'}_{16} (g_i)^2}, \qquad \sum_i w_i \overset{!}{=} 1.
\end{equation}
We tested 8$\rm^{th}$, 16$\rm^{th}$, 20$\rm^{th}$, and 24$\rm^{th}$ order Legendre polynomials and corresponding weights in an Earth atmosphere setting with the 128 bands model but did not find any significant differences  ($<$10$^{-3}$W/m$^2$) in total up/down radiative fluxes above 16$\rm^{th}$ order (not shown).  In comparison, \citet{Malik_2017} used 20$\rm^{th}$ order Gauss-Legendre polynomials after comparing to Simpson's rule integration for their radiative transfer code. Table \ref{tbl:quads} shows a comparison of the quadrature weights used in RRTM and the Gauss-Legendre weights (Eq.~\ref{eq:lq}). One can see the emphasis placed in RRTM upon probability function values close to unity, indicated by very small values for weights $i>10$, together only representing the largest $\sim$ 1$\%$ of absorption cross sections within a band. On the contrary, Gauss-Legendre weighting is symmetric around medium values of cross sections within the band, and the largest and smallest 1.3$\%$ of values are represented collectively in weights 16 and 1, respectively. The representative cross section value $\sigma_i$, or k-coefficient, corresponding to a quadrature of weight $w_i$ is then simply the arithmetic mean of the cumulative cross section function within the respective quadrature.\\
%
\subsubsection{Continuum absorption} \label{sssec:conti}
\begin{deluxetable}{c c}
    \tabletypesize{\footnotesize}
    \tablecolumns{2}
    \tablecaption{List of continua from the HITRAN CIA list\tablenotemark{a} included in REDFOX and the corresponding sources.\label{tbl:continuum}}
    \tablehead{\colhead{Molecule} & \colhead{Continuum}}    
    \startdata
         H$_2$ -- H$_2$ & \citet{abel2011} \\
         H$_2$ -- He & \citet{abel2012} \\
         CO$_2$ -- H$_2$ & \citet{wordsworth2017} \\
         CO$_2$ -- CH$_4$ & \citet{wordsworth2017} \\
         CO$_2$ -- CO$_2$ & \citet{GRUSZKA1997172} \\
         & \& \citet{baranov2003} \\
         H$_2$O $\rm_{self}$ & MT\_CKD\_3.2\tablenotemark{b} \\
         H$_2$O $\rm_{foreign}$ & MT\_CKD\_3.2\tablenotemark{b} \\
         CO$_2$ $\rm_{foreign}$ & MT\_CKD\_3.2\tablenotemark{b} \\
         N$_2$ $\rm_{mixed}$ & MT\_CKD\_3.2\tablenotemark{b} \\
         O$_2$ $\rm_{self}$ & MT\_CKD\_3.2\tablenotemark{b} \\
         O$_2$ $\rm_{foreign}$ & MT\_CKD\_3.2\tablenotemark{b} \\
         O$_3$ $\rm_{foreign}$ & MT\_CKD\_3.2\tablenotemark{b} \\
    \enddata
    \tablenotetext{a}{\url{www.hitran.org/cia/} \citep{KARMAN2019160}}
    \tablenotetext{b}{\url{http://rtweb.aer.com/continuum_frame.html} \citep{mlawer2012}}
\end{deluxetable}
Collision-induced absorption (CIA) is taken from the HITRAN CIA database \citep{KARMAN2019160}. A complete list of the currently implemented CIAs in REDFOX is shown in Table \ref{tbl:continuum}. Since the HITRAN CIA formatting is identical, any missing CIA from the list can be flexibly added in REDFOX for a specific study. Additionally REDFOX utilizes the MT\_CKD \citep{mlawer2012} version 3.2\footnote{\url{http://rtweb.aer.com/continuum_frame.html}}. Although originally developed for the H$_2$O continuum, this now also includes continua for other molecules. In REDFOX one can choose which CIA and MT\_CKD continua from Table \ref{tbl:continuum} are used. These continua are first calculated on a finer grid with 10 points per band using the same interpolation method as for  MT\_CKD in LBLRTM \citep{clough1992,clough2005}, which gives the cross sections: 
\begin{equation}
\begin{aligned}
    \sigma_i = &-\sigma_{j-1}\;\left[\frac{f_\nu}{2}(1-f_\nu)^2\right]\\
 &+\sigma_j\;\left[1-(3-2f_\nu)f_\nu^2+\frac{f_\nu^2}{2}(1-f_\nu)\right]\\
 &+\sigma_{j+1}\;\left[(3-2f_\nu)f_\nu^2 + \frac{f_\nu}{2}(1-f_\nu)^2\right]\\
 &+\sigma_{j+2}\;\left[\frac{f_\nu}{2}(1-f_\nu)\right],
\end{aligned}
\end{equation}
with 
\begin{equation}
    f_\nu = \frac{\nu_i-\nu_j}{\nu_{j+1}-\nu_j},
\end{equation}
where $i$ are the ten grid points in a band interval and $j$ are linearly interpolated from tabulated cross sections at neighboring temperatures, either from MT\_CKD or HITRAN CIA, where applicable. Values at the band boundaries are linearly interpolated from tabulated values at neighboring frequencies. We tested different numbers of points per band for the interpolation, but convergence was found for ten or more points per band.
From these interpolated cross sections,  band-integrated averages are calculated via $1/(\Delta\nu_{band}) \sum_{i=1}^n (\sigma_i\Delta\nu_i)$ and added to other opacity sources with their corresponding volume mixing ratio. \\
%
\subsubsection{UV / VIS cross sections} \label{sssec:uvvis}
In the current version of 1D-TERRA we include 81 absorbers from the MPI Mainz Spectral Atlas \citep{spectralatlas} in the visible, UV, and FUV range. These were initially chosen for the photochemistry module as described in Subsection \ref{ssec:chem}, but are also available for radiative transfer in REDFOX. Where available, we follow the recommendation of the JPL Evaluation 18 \citep{burkholder2015}. Otherwise we took the latest reference in the JPL report, unless the data coverage was poor. The full list of cross section data used in the climate and chemistry model is listed in the companion paper \citep[][in prep.]{wunderlich2019b}. These cross sections are arithmetically averaged into the corresponding REDFOX bands, and added to the other opacities discussed above. For remaining HITRAN absorbers in the band, the UV / VIS cross sections are directly added to the k-distributions for cross correlation using the random overlap approach as described earlier. In this case, UV / VIS cross sections are interpolated and band-averaged from the MPI Mainz Spectral Atlas data with the same method as described in Subsection \ref{sssec:conti} for the continua. They are used in the spectral range from the highest available energies, or wavenumbers (10$^5$ cm$^{-1}$ in the current version) up to the band with the first HITRAN data point of the given species. This approach was chosen to avoid accounting for absorption of a given species twice (i.e. both in HITRAN and MPI Mainz databases). In overlapping data regions we preferentially chose line list data rather than measured cross sections because the latter could also include experimental noise between absorption features, which would add artificial UV "continua" to our heating rates. For molecules where line list data were not included in a certain band, the cross sections are simply added as mean opacities per band in the same way as the continua are added in Subsection \ref{sssec:conti}.
%
\subsubsection{Rayleigh Scattering} \label{sssec:rayleigh}
\begin{deluxetable}{c c c c}
    \tabletypesize{\footnotesize}
    \tablecolumns{4}
    \tablecaption{List of parameters for the Rayleigh cross section calculations. A and B from \citet{allen1973} and \citet{VARDAVAS1984} are used for the refractivity approximation, D \citep{sneep2005} is for depolarisation in the "King Factor" K$_\lambda$. \label{tbl:depol}}
    \tablehead{\colhead{Molecule} & \colhead{A} & \colhead{B} & \colhead{D}}    
    \startdata
        CO  & 32.7  & 8.1   & 9.49$\cdot$ 10$^{-3}$ \\
        CO$_2$  & 43.9  & 6.4   & 8.05$\cdot$ 10$^{-2}$ \\
        H$_2$O  & --  & --   & 3.00$\cdot$ 10$^{-4}$ \tablenotemark{a} \\
        N$_2$  & 29.06  & 7.7   & 3.05$\cdot$ 10$^{-2}$ \\
        O$_2$  & 26.63  & 5.07   & 5.40$\cdot$ 10$^{-2}$
    \enddata
    \tablenotetext{a}{\citet{murphy1977}}
\end{deluxetable}
\begin{deluxetable}{c c c}
    \tabletypesize{\footnotesize}
    \tablecolumns{3}
    \tablecaption{List of measured reference Rayleigh cross sections $\sigma_{0,i}$ at wavelength $\lambda_{0,i}$ from \citet{shardanand1977}.\label{tbl:sigrayref}}
    \tablehead{\colhead{Molecule} & \colhead{$\sigma_{0,i}$ [cm$^2$]} & \colhead{$\lambda_{0,i}$ [$\mu$m]}}    
    \startdata
        H$_2$   & 1.17 $\cdot$ 10$^{-27}$  & 0.5145 \\
        He   & 8.6 $\cdot$ 10$^{-29}$  & 0.5145 \\
        CH$_4$   & 1.244 $\cdot$ 10$^{-26}$  & 0.5145 \\
    \enddata
\end{deluxetable}
Rayleigh scattering is included in our model \citep[see also][]{vonparis2015} for CH$_4$, CO, CO$_2$, H$_2$, H$_2$O, He, N$_2$, and O$_2$. The Rayleigh scattering cross sections  (cm$^{2}$) for CO, CO$_2$, H$_2$O, N$_2$, and O$_2$ are calculated using \citep{allen1973} : 
\begin{equation}\label{eq:r0}
   \sigma_{\rm{Ray},i}(\lambda) = \frac{32\pi^3}{3\;n^2}\;\frac{r_i^2\;K_{\lambda,i}}{\lambda^4}, 
\end{equation}
and with number density $n=2.688\cdot10^{19}$~cm$^{-3}$ for standard temperature and pressure (STP):
\begin{equation}\label{eq:r1}
   \sigma_{\rm{Ray},i}(\lambda) = 4.577\cdot 10^{-21}\;\frac{r_i^2\;K_{\lambda,i}}{\lambda^4}, 
\end{equation}
where  $\lambda$ is in $\mu$m, $r_i$ represents the  refractive index $(n-1)$ of species $i$, and the "King factor" $K_{\lambda,i}$ is a correction for polarization, given by:
\begin{equation}\label{eq:r2}
    K_{\lambda,i} = \frac{6+3D_i}{6-7D_i},
\end{equation}
with $D_i$ being the depolarization factor calculated from \citet{sneep2005} shown in Table \ref{tbl:depol}. Further, for CO, CO$_2$, N$_2$, and O$_2$, the  refractive index $r$ is approximated using \citep{allen1973}:
\begin{equation}\label{eq:r3}
    r_i = 10^{-5}\;A_i\;\left(1+\frac{10^{-3}\;B_i}{\lambda^2} \right),
\end{equation}
with the two additional parameters $A_i$ and $B_i$, also shown in Table \ref{tbl:depol}, and $\lambda$ again in $\mu$m. For H$_2$O the  refractive index is calculated as $r_{\rm H_2O}$ = 0.85 $r_{\rm dryair}$ \citep{edlen1966}, with the approximation for the  refractive index of dry air \citep{bucholtz1995}:
\begin{equation}\label{eq:r4}
    r_{\rm dryair} = 10^{-8}\;\left(\frac{5.7918\cdot 10^{6}}{2.38\cdot 10^{2} - \lambda^{-2}} + \frac{1.679\cdot 10^{5}}{57.362 - \lambda^{-2}}\right).
\end{equation}\\
Rayleigh scattering cross sections for H$_2$, He, and CH$_4$ are included using reference measurements of $\sigma_{0,i}$ at wavelengths $\lambda_{0,i}$ from \citet{shardanand1977} shown in Table \ref{tbl:sigrayref}. These cross sections are then approximated for other wavelengths using the simple $\lambda^{-4}$ relationship:
\begin{equation}\label{eq:r5}
    \sigma_{\rm{Ray},i}(\lambda) = \sigma_{0,i}\;\left(\frac{\lambda_{0,i}}{\lambda}\right)^4,
\end{equation}
where $\sigma_{\rm{Ray},i}$ of species $i$ are given in cm$^2$/molecule, $\lambda$ is the wavelength of interest in $\mu$m, and $\lambda_{0,i}$ is a reference wavelength where $\sigma_{0,i}$ was measured.
 In REDFOX we updated the calculation of band- integrated scattering cross sections, which is now performed for all spectral bands, rather than only the SW bands. The band-mean values are now calculated from one specific wavelength per band, representative of the  band-integrated average. We emphasize that this does not correspond to the band-mid wavelength. Instead we use the equivalence  of the integrals:
 \begin{equation}\label{eq:r6}
 \begin{aligned}
    \int_{\lambda_1}^{\lambda_2} \lambda^{-4} d\lambda \quad&\overset{!}{=}\quad \int_{\lambda_1}^{\lambda_2} \lambda_{\rm ray}^{-4} d\lambda \quad = \quad\lambda_{\rm ray}^{-4}\int_{\lambda_1}^{\lambda_2}d\lambda\\
    \Rightarrow{} \lambda_{ray} &= \left(\frac{\lambda_1^{-3}-\lambda_2^{-3}}{3\,(\lambda_2-\lambda_1)}\right)^{-\frac{1}{4}} ,
 \end{aligned}
 \end{equation}
 to determine in each band the representative wavelength $\lambda_{ray}$ for the Rayleigh cross section calculations ($\lambda$ in equations \ref{eq:r1}, \ref{eq:r3}, \ref{eq:r4}, and \ref{eq:r5}). Then the choice of  the radiative transfer method (see Subsection \ref{sssec:opacities}) defines in which bands the Rayleigh scattering opacities are used.
%
\subsubsection{Source Functions} \label{sssec:sources}
We consider essentially two radiative sources, incident stellar flux and thermal radiation from the planetary surface and the atmosphere. Stellar spectra can be either taken directly from the MUSCLES database\footnote{\url{archive.stsci.edu/prepds/muscles/}}, or from the VPL webpage\footnote{\url{vpl.astro.washington.edu/spectra/stellar/}}. The solar spectrum is taken from \citet{gueymard2004}. Spectra which do not cover the entire wavelength range, were extended with the NextGen 4 spectrum of the corresponding effective temperature of the star up to 971~$\mu$m \citep{hauschildt1999}. We first perform a logarithmic interpolation of the spectra to a constant resolution of 1~$\rm\AA$ over the entire wavelength range and then bin the data into our 128 bands using an arithmetic mean.\\
Thermal black body emissions from the planetary surface and every atmospheric layer are calculated for 101 points (100 intervals) per band distributed equally in $\nu$ following Planck's law in wavenumbers,
\begin{equation}
    B_\nu(\nu,T) = 2hc^2\nu^3\;\frac{1}{e^{(\frac{hc\nu}{k_BT})}-1}.
\end{equation}
Then a simple arithmetic mean is taken as the average black body source term in the given band and atmospheric layer. We tested this simple method against the computationally slightly more expensive trapezoidal rule in the Earth and Venus atmosphere runs but found no significant ($<10^{-4}$) differences in cooling rates or atmospheric temperature structure. This suggests that our band model features sufficiently small band widths. For this reason we chose the faster band mean calculations explained above.
%
\subsubsection{Opacities and Transmission Functions} \label{sssec:opacities}
Once all individual k-coefficients $\sigma_i$ are calculated, they are mixed for the atmosphere of interest. In random overlap, $\sigma_{ij,mn,b}$ in units of cm$^2$/molecule of two molecules $m$ and $n$ and their respective quadratures $i$ and $j$ in band $b$ are cross correlated when performing radiative transfer calculations by: 
\begin{equation}\label{eq:ro}
    \sigma_{ij,mn,b} = \frac{\sigma_{i,m}\chi_{m} + \sigma_{j,n}\chi_n}{\chi_{m}+\chi_n},
\end{equation}
where $\chi$ is the respective volume mixing ratio, and individual k-coefficients $\sigma$ are linearly interpolated in $T$ and log($p$) from the pre-calculated k-distributions. The corresponding 16x16 quadrature weights as a result of mixing the k-coefficients from two molecules, are:
\begin{equation}
    w_{ij} = w_i w_j, 
\end{equation}
with $i$ and $j$ each taking integer values from 1-16 as shown in Table \ref{tbl:quads}. In REDFOX we use $quicksort$ \citep{nr} to sort the mixed $\sigma_{ij}$ and corresponding $w_{ij}$ again as a monotonically increasing function, or k-distribution. Before adding new molecules to the mix, we bin the mixed k-distribution back into the original 16 quadratures ($g$=1-16). For the respective mean k-coefficients $\sigma_{g}$, individual $\sigma_{ij}$ are added where the cumulative weights $w_{c,ij}$ of the mix lie in the range $w_{c,i-1} < w_{c,ij} < w_{c,i}$ of the original quadrature with weight $w_i$. The new k-coefficients are then calculated as: 
\begin{equation}
    \sigma_{g,mn} = \frac{1}{w_i}\sum_{w_{c,ij}=w_{c,i-1}}^{w_{c,ij}=w_{c,i}} \left(\frac{w_{ij}}{w_i}\right)\; \sigma_{ij,mn},
\end{equation}
with 
\begin{equation}
    w_{c,i} = \sum_{i=1}^{g} w_{i} \quad and \quad w_{c,ij} = \sum_{ij=1}^{g^2} w_{ij}.
\end{equation}
Contributions of mix-quadratures that overlap with the boundaries of the original 16 quadratures, are factored into the corresponding $\sigma_g$ accordingly. This process is repeated for every molecule added to the gas mixture. Depending on the composition of the atmosphere of interest, one cannot expect that all constituents are accounted for by our absorbers, e.g. when we choose to exclude specific constituents from radiative transfer calculations e.g. for performance reasons, studies of individual gas contributions, or when e.g. noble gases are excluded in calculations because they would not contribute significantly to the radiative energy budget. Since Eq. \ref{eq:ro} assumes the absorbing contribution of all constituents, we calculate the total mixing ratio of the absorbing species $\chi_{\rm mix}$ along with the k-coefficients, and correct the cross sections of the mixture using $\sigma_g$\,=\,$\sigma_g\cdot\chi_{\rm mix}$. This approach is especially appropriate for Earth's atmosphere, where N$_2$ accounts for approx. 78\% of molecules, while not contributing significantly to absorption. In this case calculations can be sped up significantly when we exclude non-significantly-absorbing species from the radiative transfer calculations. \\
Once opacities, $\tau\,=\,\sigma_g u$, with $u$ being the respective column density, are calculated, other previously described band- integrated opacities can be added flexibly.  Radiative transfer is solved together with Rayleigh scattering, for which we use the  quadrature $\delta$-two-stream approximation \citep{meador1980,toon1989} for SW radiative transfer. For the thermal outgoing LW radiative transfer we have the option to use either the \citet{toon1989}  hemispheric-mean two-stream approximation including Rayleigh scattering and with Planck emissions instead of the stellar component, or the faster diffusivity approximation \citep{mlawer1997} without scattering. In this case, the radiative transfer equation is solved for the single angular point $\mu=\frac{1}{1.66}$, corresponding to  a polar angle $\theta$=52.95$^\circ$ \citep{elsasser1942}, and the linear-in-tau approach for the Planck functions \citep{clough1992}.  While the implementations of the above-mentioned radiative transfer solvers date back to \citet{pavlov2000} and \citet{segura2003}, they are still useful in our cloud-free studies. Upcoming model updates including scattering by larger aerosols and cloud particles will include two-stream updates similar to \citet{heng2018} and an option to use a discrete ordinates method, such as DISORT \citep[e.g.][]{stamnes1988,hamre2013}, which will then allow for treatment of incident stellar beam, scattering, and the internal source function, namely the Planck function of each layer, at the same time, ridding us of the SW - LW distinction necessary with the currently used two-stream approximations.\\
The globally-averaged zenith angle ($\theta$) for  effective path lengths  of the direct stellar beam through an atmospheric layer can be chosen by the user. \citet{cronin2014} has pointed out that zenith angle and global-mean averaged irradiation ($S$) have to be changed together according to the product $S\cdot\cos{\theta}$ = $S_0$/4, where $S_0$ is the Total Stellar Irradiation (TSI), and 1/4 comes from the purely geometric derivation of planetary cross section over planetary surface area.  Note that a simple daytime-weighted global mean assumption, a zenith angle $\theta=60^\circ$ with $S=S_0/2$ is commonly used in the literature. Additionally this approximation assumes a rapidly rotating planet, such that the incident stellar flux is effectively distributed over the entire planetary surface despite only illuminating one half of the planet. This would obviously be different for tidally-locked planets, where a more general parameter can be introduced that simulates the efficiency of heat redistribution from the permanent day to the night side \citep[see, e.g., ][]{spiegel2010,Malik_2017}.
\\
While aerosols and clouds can be treated to some degree in the two-stream approximation via extinction coefficients which depend on particle sizes, forms and refractive indices \citep[see, e.g.,][]{kitzmann2010}, we do not explicitly include such particles in the current work. We parameterize cloud behavior in a simple way by increasing the surface albedo and adjusting the effective path length in radiative transfer with a zenith angle deviating from the geometric mean value of 60$^\circ$.  The appropriate choice of albedo and zenith angle, however, is in general a degenerate problem.\\
 With all the model updates mentioned above, calculation times are nevertheless sufficient for our purposes. While calculating opacities plus radiative fluxes in MRAC for 4 absorbers in 25 LW bands and 38 SW bands plus Rayleigh scattering took $\sim$180~milliseconds, REDFOX with the same configuration takes $\sim$235~milliseconds. With our long list of LW and SW absorbers in our 128 bands including MT\_CKD and CIAs now takes $\sim$1600~milliseconds, all on a single CPU. This is still fast compared to full line-by-line calculations, which can take several minutes on a server node of 20 CPUs.

\subsubsection{Key Features of REDFOX} \label{sssec:features}
The following is a summary of the key features of REDFOX in the configuration validated and applied in the current work:
\begin{itemize}
    \item continuous k-distribution model with 128 spectral bands (16 $g$ Gauss-Legendre)
    \item Spectral range $\nu = $ 10$^5$ cm$^{-1}$ (100~nm) - 0 cm$^{-1}$ ($\infty$)
    \item Pressure range 0.01~Pa - 10$^3$~bar
    \item Temperature range 100~K - 1000~K
    \item 20 absorbers (HITRAN 2016) -- see Table~\ref{tbl:molecules}
    \item 81 VIS/UV/FUV absorbers (MPI Mainz spectral atlas) -- see Table~\ref{tbl:molecules}
    \item CIAs (HITRAN CIA list) -- see Table~\ref{tbl:continuum}
    \item MTCKD\_3.2 -- see Table~\ref{tbl:continuum}
    \item flexible choice of absorbers (SW/LW, CIA/CKD)
    \item flexible choice of SW/LW ranges for specific study
\end{itemize}

\subsection{Climate Module} \label{ssec:clima}
%

Two central tasks of the climate module are to calculate atmospheric and surface temperatures in radiative-convective equilibrium, and to calculate temperature-dependent molecular abundances, of condensible species e.g. H$_2$O or CO$_2$ in convective regions where temperatures change. The governing energy-conserving  differential equation determining the temperature $T$ profiles is \citep[see e.g.][]{pavlov2000}:
\begin{equation}\label{eq:dt}
    \frac{d}{dt}T(z) = -\frac{g(z)}{c_p(T,z)}\frac{dF(z)}{dp(z)},
\end{equation}
where $d/dt$ is the  time derivative, $g(z)$ is gravitational acceleration, $c_p$ the heat capacity of the atmospheric layer, $F$ is the radiative flux, and $p$ the cell-centered pressure of the atmospheric layer at mid-layer height $z$.  Eq.~\ref{eq:dt} is discretized in the model using finite differences, resulting in $dt$ being used as the finite time step. After the radiative temperature profile is calculated for a model time-step, radiative lapse rates are compared in every layer with adiabatic lapse rates to determine where convective adjustment has to be performed (i.e. Schwarzschild criterion). The decision as to the appropriate adiabatic lapse rate (moist or dry) and whether H$_2$O or CO$_2$ is the condensing molecule in the moist case, is based on the fraction of respective partial pressures over the corresponding saturation vapor pressure.  Note that combined moist adiabatic lapse rates are not implemented. If CO$_2$ condensation occurs, the resulting lapse rate is a CO$_2$ moist adiabat, disregarding any possible H$_2$O contribution. H$_2$O moist adiabatic lapse rates are calculated if H$_2$O condenses in the absence of CO$_2$ condensation. H$_2$O profiles are calculated using prescribed relative humidity (RH) profiles e.g. by \citet{manabe1967}, for planetary scenarios with a water reservoir, i.e. an ocean, via:
\begin{equation}\label{eq:h2ocalc}
    \chi_{\rm H_2O}(p) = \rm{RH}(p)\,\frac{p_{\rm sat,H_2O}(T(p))}{p},
\end{equation}
with $p_{\rm sat}$ being the saturation vapor pressure. 
The total surface pressure $p_0$ of the atmosphere is calculated via $p_{0}\,=\,p_{0,\rm dry}+p_{0,\rm H_2O}$; therefore atmospheric pressures,  and with them the mixing-ratios of all atmospheric constituents, are adapted for those convective layers. For a detailed description of convective adjustment  and CO$_2$ lapse rates in our model we refer to e.g. \citet{kasting1991,kasting1993}, as well as to \citet{vonParis2008} for the water vapor treatment.\\
Heat capacities ($c_p$) for temperature and lapse rate calculations have  historically been included  in our model for different species using different approaches \citep[for details see e.g.][]{vonParis2008,vonParis2010,vonparis2015}.  The next 1D-TERRA update will come with updated $c_p$ treatment. Therefore we only give a brief summary of our currently used $c_p$ calculations in Appendix A. 
%
\subsection{Photochemistry Module} \label{ssec:chem}
The atmospheric composition profile of each species is calculated in the photochemistry module using the continuity-transport equation:
\begin{equation}\label{eq:conteq}
   \frac{\partial n}{\partial t} = \frac{\partial}{\partial z} \left(K \cdot \frac{\partial n}{\partial z} \right)  + P -  n L ,
\end{equation}
where $n$ denotes the number density of a certain species (cm$^{-3}$), $P$ the production term (cm$^{-3}$~s$^{-1}$), $L$ is the loss rate (s$^{-1}$), and $K$ the eddy diffusion coefficient (cm$^{2}$~s$^{-1}$). The loss and production rates are determined by the kinetic coefficients and the boundary conditions. The companion paper of \citet[][in prep.]{wunderlich2019b} describes the new photochemistry module BLACKWOLF with a flexible chemical network. The full network consists of  1127 reactions for 115 species including photolysis for 81 absorbers. \\
The scheme can consider wet and dry deposition, as well as an upward flux at the surface for each species. Outgassing from volcanoes is treated as a upward flux equally distributed over the lower 10~km of the atmosphere. At the top-of-atmosphere the module can calculate escape fluxes and parameterized effusion flux, or use predefined values. Usually we only consider O, H and H$_2$ to have an upper boundary flux. The profile of $K$ can be either a fixed predefined profile or is calculated in the model depending on the planetary atmosphere. See \citet[][in prep.]{wunderlich2019b} for the calculation of $K$ and more details on the photochemical module.
%
\section{Model Validation} \label{sec:valid}
%
\picwide{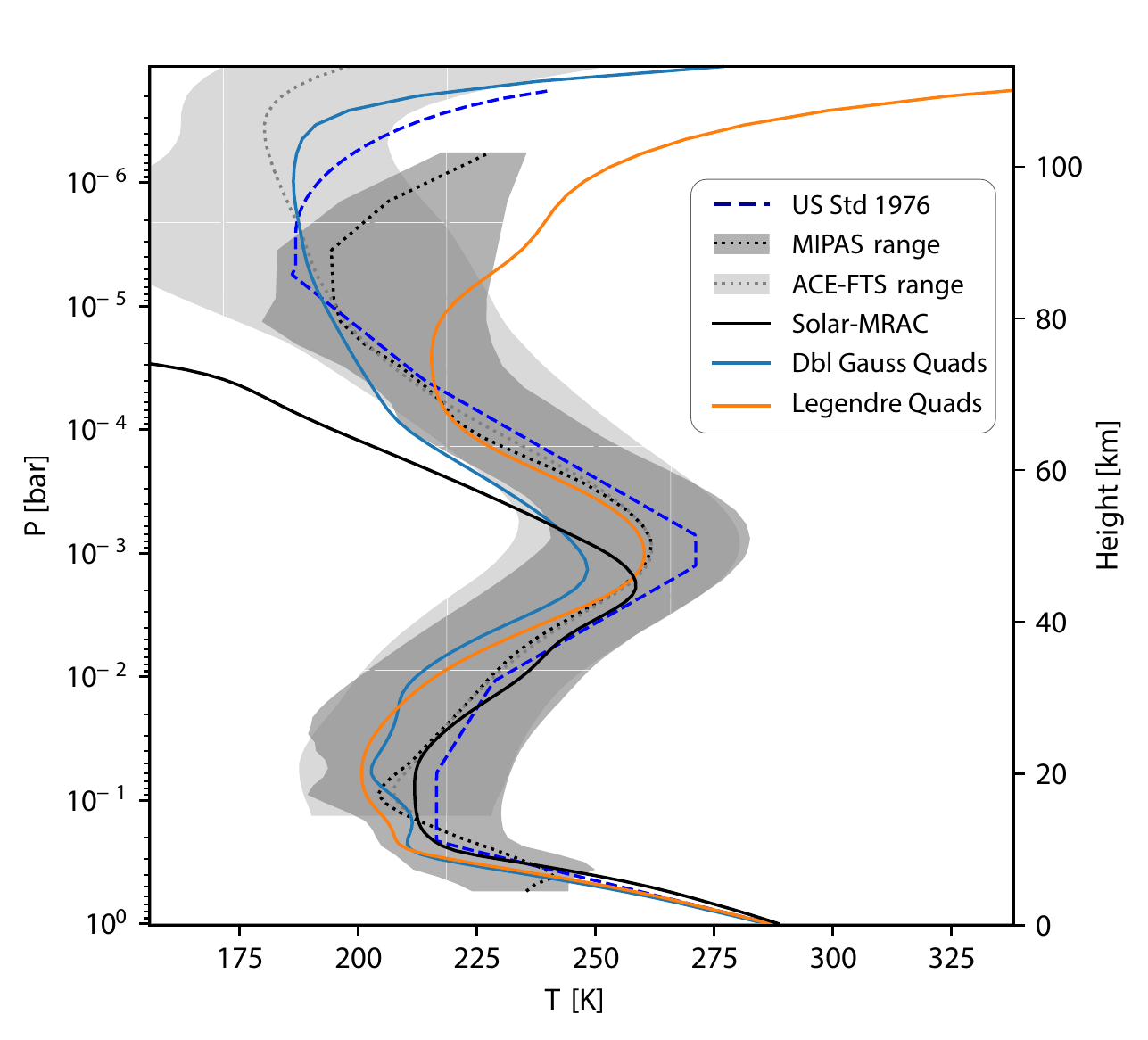}{Temperature profiles for Earth's modern atmosphere. We compare two versions of REDFOX with k-distributions calculated with Legendre (orange) and double Gaussian (blue) quadratures against the US standard atmosphere (blue-dashed) and our original model MRAC (black). Overplotted are 2$\sigma$ ranges of Earth observation data from MIPAS and ACE-FTS.}
\picfullwide{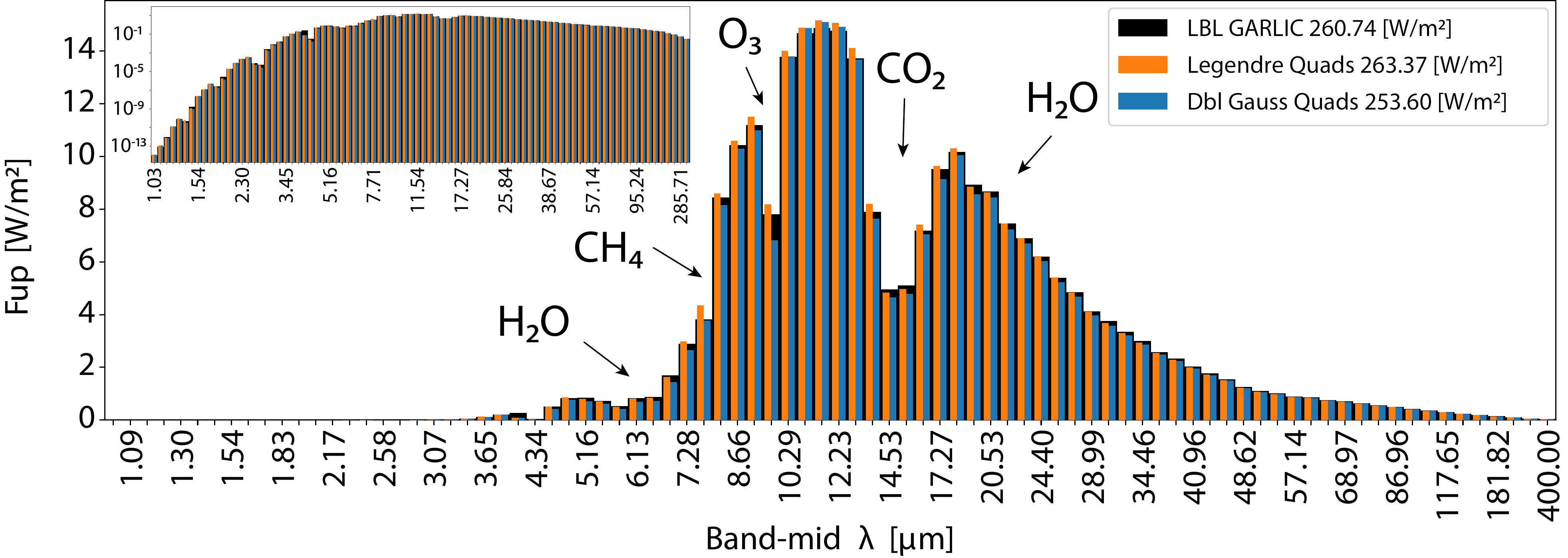}{Earth validation of OLR fluxes against the LBL code GARLIC (black). The small insert shows fluxes in logarithmic scaling. We compare both quadrature settings in REDFOX (orange and blue) and indicate regions with major absorbers with arrows.}
\picwide{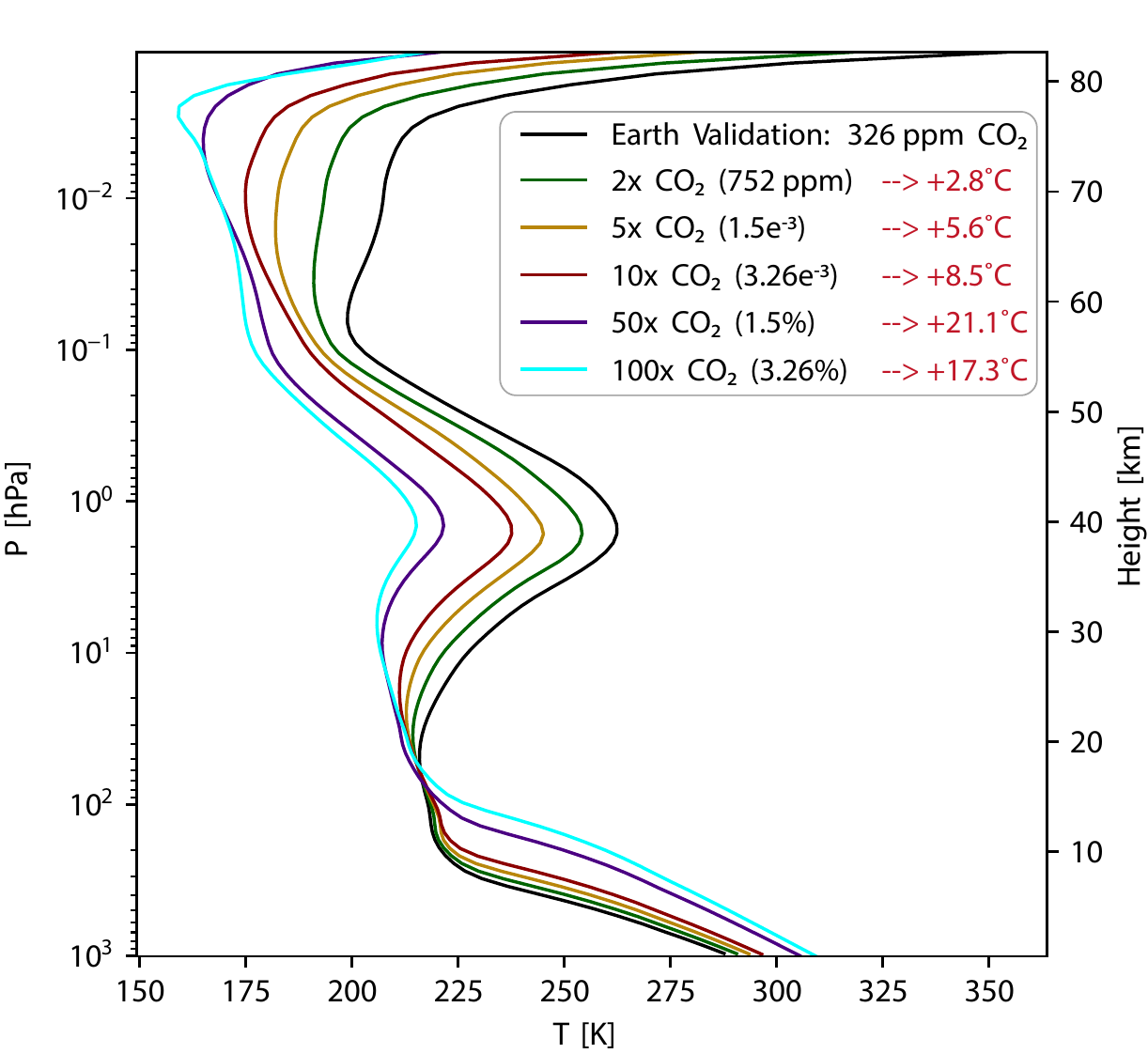}{Earth climate repsonse study for increased amounts of CO$_2$. The red values in the legend represent the resulting increase in surface temperatures.}
\picfull{Steam}{Comparison of radiative fluxes under steam atmosphere conditions compared against the tested models in \citet{yang2016}.}
\picwide{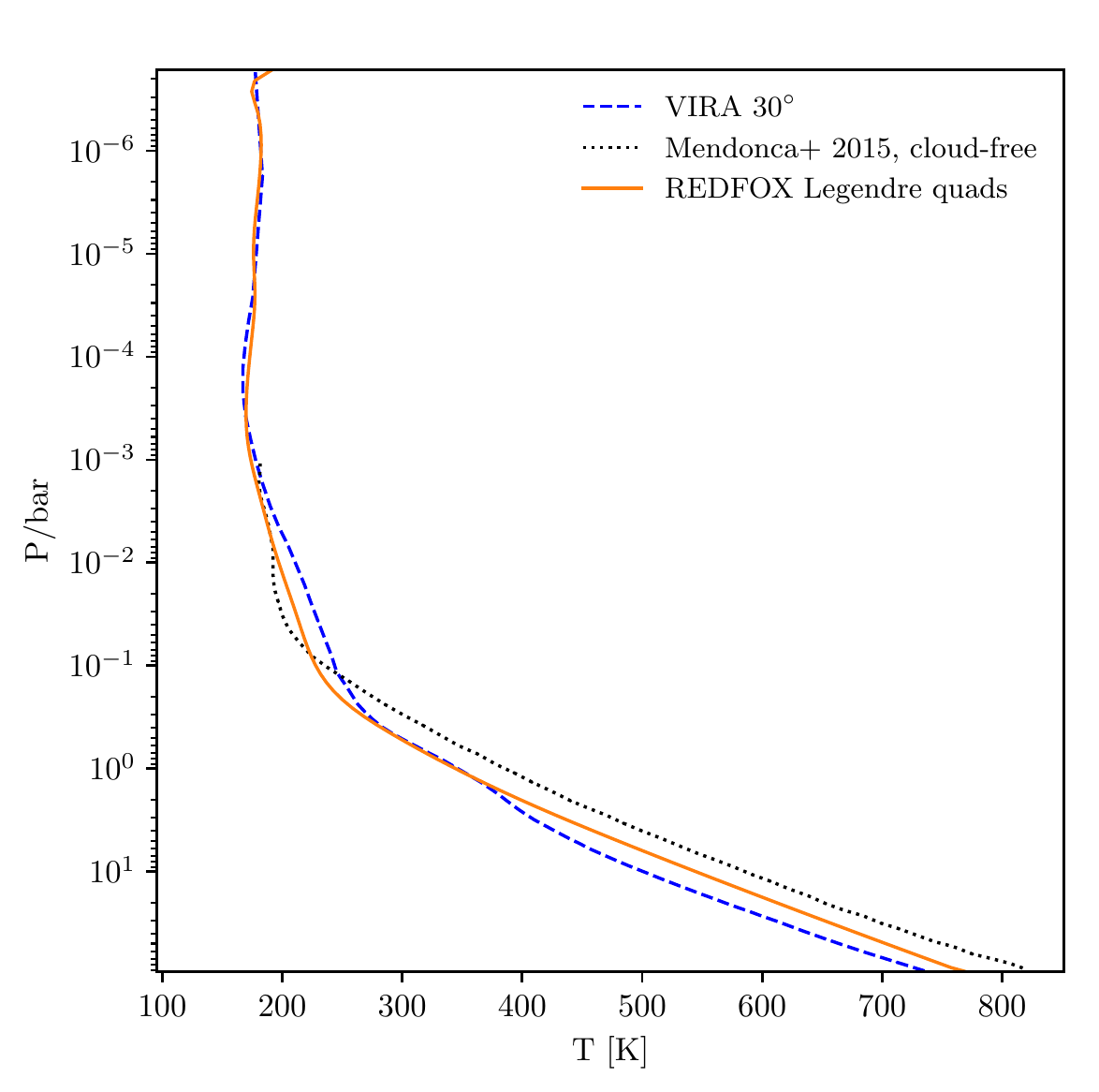}{Validation for cloud-free Venus. We compare our climate model (orange) to the Venus reference atmosphere (blue-dashed) and the cloud-free model calculations from \citet{mendonca2015} (black-dotted).}
\picwide{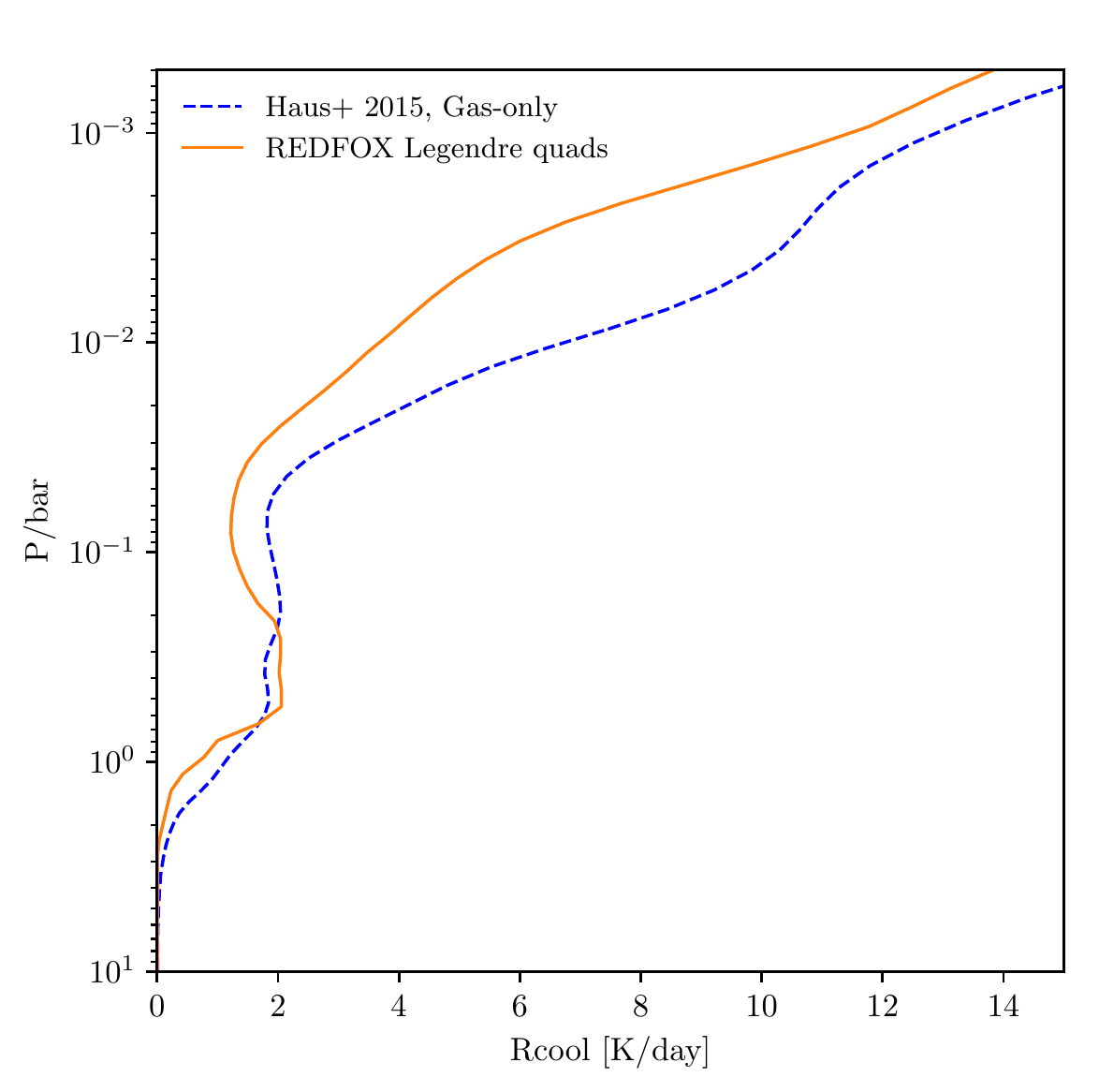}{Cooling rates in our cloud-free Venus model atmosphere with REDFOX (orange) compared to gas contributions to the cooling rate in the cloudy VIRA atmosphere taken from \citet{haus2015} (blue).}
\picfullwide{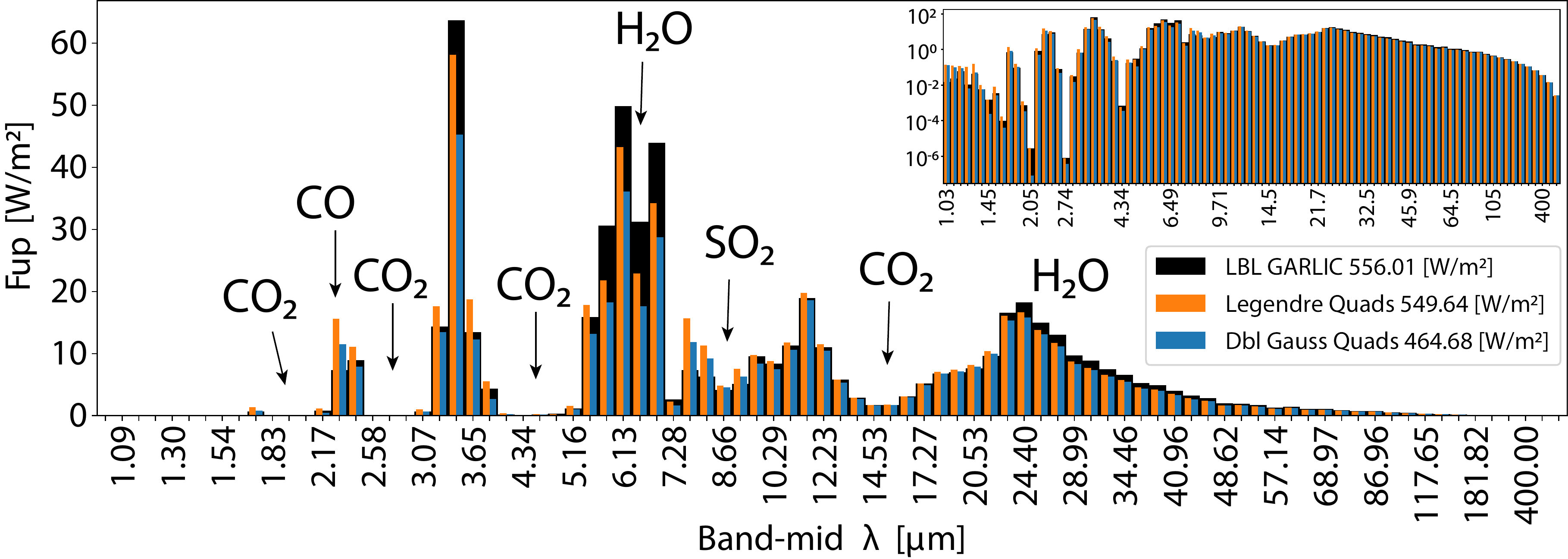}{As for Figure \ref{Earth-valid-fluxes} but for cloud-free modern Venus conditions.}
\picwide{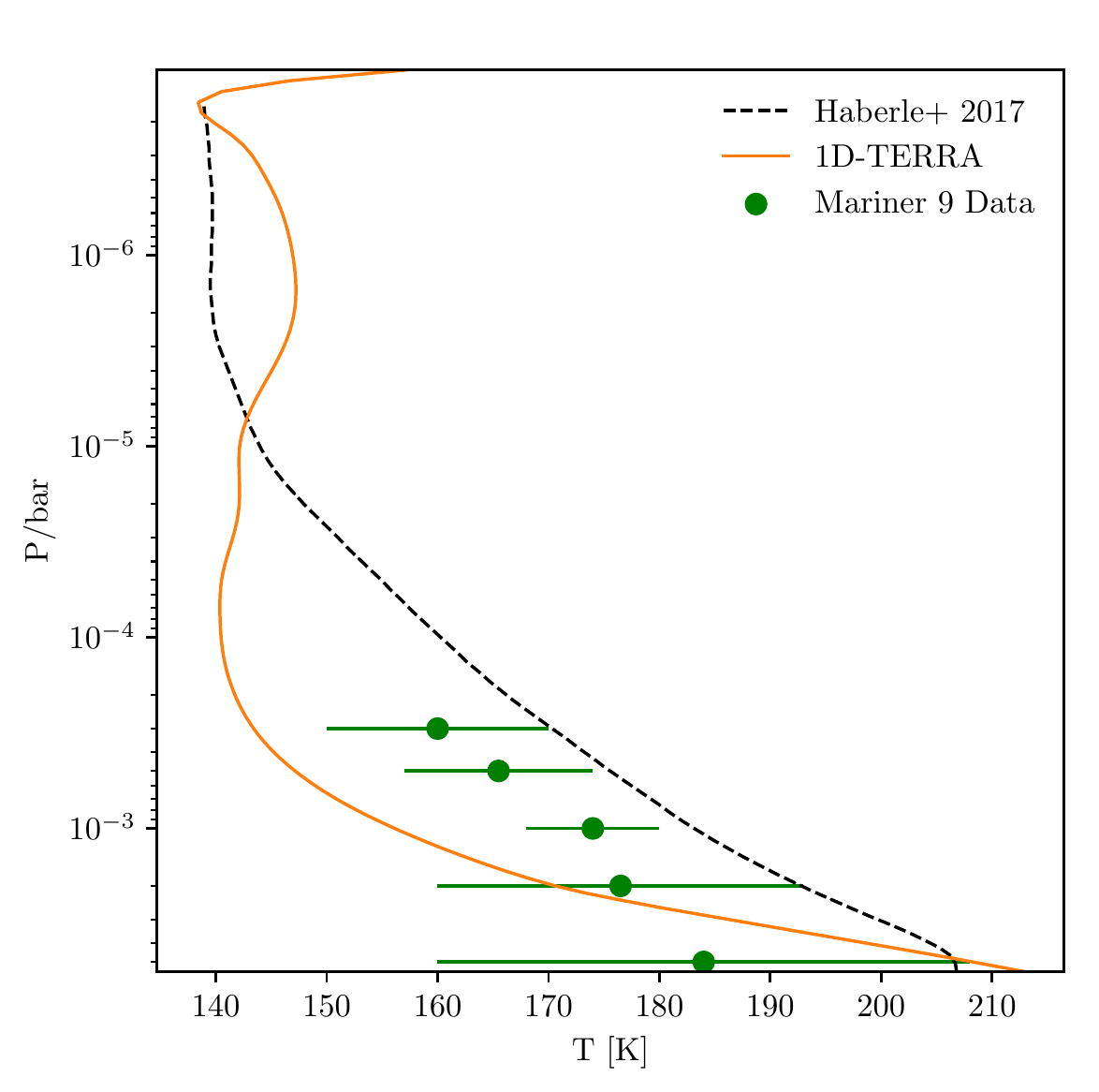}{Validation for a hypothetically aerosol-free Mars. We compare temperatures calculated with 1D-TERRA (orange) to the \citet{haberle2017} Mars reference atmosphere (black) and Mariner 9 IRIS data (green).}
In this section we restrict ourselves to the validation of the radiative transfer module, REDFOX, and its integration into the climate part of 1D-TERRA, thus in some cases we prescribe the atmospheric composition. The chemistry is validated separately in the companion paper \citet[][in prep.]{wunderlich2019b}. We first validate for Earth standard conditions, then go on to Earth climate-change studies with increased amounts of CO$_2$, before validating for other Solar System planets, i.e. modern Venus and Mars, and then exoplanetary scenarios (i.e. Earth-like and steam atmospheres). In all cases we discretize the atmosphere into 100 layers from the surface up to the top-of-atmosphere (TOA).
%
\subsection{Modern Earth}
Figure \ref{Earth-Temp} shows our 1D-TERRA results in coupled climate-chemistry mode for US standard 1976 Earth atmospheric composition.  We tune surface albedo and zenith angle for radiative transfer calculations together to achieve Earth global average surface temperatures of 288.15~K in our cloud-free model and an Earth-like global mean O$_3$ column in the range 300-320 Dobson Units (DU) \citep[see e.g.][]{segura2010}, which is achieved with a surface albedo of 0.255 and an angle $\theta$~=~54.50$^\circ$, thus $S$~=~0.4305~$S_0$ so that $S\cdot\cos(\theta)$~=~$S_0$/4, for all radiative transfer calculations in climate and chemistry. This agrees well with the fact that modern Earth is partially cloudy; this angle lies well in the range for a clear sky and a fully cloud-covered planet \citep[see~e.g.][]{cronin2014}.   Note that, zenith angles used in climate and chemistry calculations do not necessarily have to be the same, see, e.g., \citet{Hu2012} for a discussion. Unless stated otherwise we will use this parameter set for all other tests and applications. \\
Temperature profiles with the original radiative transfer model MRAC \citep{vonparis2015} were calculated using absorption by H$_2$O, CO$_2$, CH$_4$, and O$_3$, with the addition of O$_2$ in one band (752-784 nm) for stellar absorption. Those with REDFOX are calculated with all 20 HITRAN absorbers from Table \ref{tbl:molecules} with the addition of UV/VIS cross sections for CH$_3$Cl, CH$_4$, CO, CO$_2$, H$_2$O, HNO$_3$, N$_2$, N$_2$O, NO, O$_2$, and O$_3$, for wavenumbers above available HITRAN data, as described in Subsection \ref{sssec:uvvis}. We calculate temperature profiles with two different k-distribution quadrature settings in REDFOX as shown in Table \ref{tbl:quads} to determine which one compares better to observations. For this we compare to the US standard atmosphere 1976, as well as Earth observation data from MIPAS\footnote{\url{www.imk-asf.kit.edu/english/308.php}} \citep{clarmann2009} and ACE-FTS\footnote{\url{ace.scisat.ca/publications/}} \citep{boone2005}. From both observation data sets, we show (grey shaded) the 95\% (2$\sigma$) range and the global annual mean temperature profiles (dotted lines). While we see generally good agreement between our calculated low and mid atmospheric temperature profiles with the US standard atmosphere and the MIPAS and ACE-FTS data (see Fig. \ref{Earth-Temp}), in all model runs using the double Gaussian quadratures, both updated (blue solid) and original model (black solid), the stratopause peak lies slightly below US standard atmospheric values in both, height and temperature, similar to other cloud-free 1D studies \citep[see, e.g.,][]{segura2003}. Mesospheric temperatures in these cases lie slightly below the MIPAS and ACE-FTS 2$\sigma$ ranges. Figure~\ref{Earth-Temp} suggests that Gauss-Legendre quadratures (orange) result in a temperature profile closer to US standard atmospheric values and global mean averages from observations up to $\sim$70~km. The inclusion of the Legendre-Gauss quadratures therefore represents a major improvement in REDFOX. \\
Additionally, with the addition of O$_2$ absorption cross sections in the UV Schumann-Runge bands, the model now clearly shows mesopause and thermospheric temperature inversions that were previously lacking (in MRAC). In this range, calculations with REDFOX using the double Gaussian quadratures compare better with the observations. However, we are not taking into account any UV absorption in the atmosphere above our TOA, hence our O$_2$ absorption might be slightly overestimated, leading to generally warmer mesopause values, especially visible in the calculations with the Gauss-Legendre quadratures. In previous studies with our 1D climate-chemistry model the TOA was set to $\sim$6~Pa due to the lack of O$_2$ absorption in the UV, similar to e.g. \citet{segura2010,kopparapu2013,ramirez2014}, while other studies \citep[e.g.][]{meadows2018} have set isothermal profiles above an arbitrary model pressure where model temperatures would otherwise strongly underestimate temperatures due to the missing physics. \\
Figure \ref{Earth-valid-fluxes} shows the model TOA outgoing long-wave radiation (OLR) for the two previously discussed different quadrature weightings (orange and blue) compared to LBL calculations with GARLIC (black).  We use GARLIC with HITRAN 2016 and MT\_CKD for H$_2$O continuum, as in REDFOX. The spectral grid in GARLIC is discretized with $\delta\nu=\gamma/4$ and the Voigt profiles are cut at 25~cm$^{-1}$. The TOA OLR calculated by GARLIC is 260.74~W/m$^2$. Results for REDFOX using Gauss-Legendre quadratures ( 263.37~W/m$^2$) show very good agreement within  1.0$\%$, while the double Gaussian quadratures (253.60~W/m$^2$) differ by  2.7$\%$. Figure~\ref{Earth-valid-fluxes} suggests that fluxes estimated with the Gauss-Legendre method (orange columns) compare generally better with the LBL fluxes (black columns).  \citet{kopparapu2013} validated their k-distribution model, using the same double Gaussian quadratures, against a different LBL code SMART \citep{meadows1996} for different terrestrial atmospheres, with deviations of the k-distribution model in the range between 2.7~-~4.0\%, similar to our double Gaussian validation. Note that RRTM \citep{mlawer1997}, using the double Gaussian quadratures, was validated for Earth conditions to deviate only $<$~1W/m$^2$, when compared to the same LBL model, namely LBLRTM, that was used to derive the k-distributions in the first place. Only this way one can be certain to not mix in the differences between LBL models, see, e.g., \citet{schreier2018agk,schreier2018ace} for a LBL model comparison. Based on this comparison we now proceed using the Gauss-Legendre quadratures for the remaining validations and case studies.
%
\subsection{Earth's Climate response to CO$_2$ increase}
Here we test our modeled climate response to increased levels of CO$_2$, and compare to results from studies of climate change on Earth. \citet{rogelj2012} summarized results from an extensive list of 3D model studies and the International Panel on Climate Change (IPCC) report estimates for a doubling of CO$_2$ levels from the 1976 values of 326~ppm to 652~ppm. Their findings suggest a resulting change in global mean surface temperatures in the range from +2.6 to +3.6$^{\circ}$C. Upon doubling CO$_2$, our resulting change in global mean surface temperatures of +2.8$^{\circ}$C agrees very well with these works. Figure \ref{Earth-2xco2} presents our results for the 1976 CO$_2$ value (black) and gradually increased levels of CO$_2$ by factors of 2, 5, 10, and 100.  We further see the typical behavior of CO$_2$ as a greenhouse gas, confining thermal radiation in the lower atmosphere, thus heating the troposphere while cooling the middle and upper atmosphere.
%
\subsection{Steam Atmospheres}
To further test H$_2$O absorption in REDFOX, we compute radiative fluxes output from REDFOX for the scenarios of the model inter-comparison study by \citet{yang2016} with fixed temperatures and atmospheric compositions. We took their values for Earth-like planets having a 1~bar N$_2$ atmosphere with 376~ppm CO$_2$ and H$_2$O steam that varies as a response to global mean surface temperatures between 250-360~K at saturation. With these added amounts of steam the surface pressure changes accordingly  (up to +0.5~bar), representing evaporation from oceans. Temperature profiles are set to 200~K isotherms above the tropopause. The study uses Earth-like global mean stellar irradiation of 340~W/m$^2$ for all tests on specifying the above-mentioned surface temperatures, thus, OLRs do not balance incoming radiation. We specify their boundary parameters with REDFOX and plot output from both studies for different LBL and k-distribution models as shown in Figure~\ref{Steam}. REDFOX compares well with the models shown, and suggests the symptomatic OLR flattening for higher temperatures towards the Kobayashi-Ingersoll limit \citep[see, e.g.,][]{nakajima1992} (lower-right), as well as the transition from optically thin to optically thick for thermal radiation in the first atmospheric layer above the surface due to the increased steam (lower-left).  Note that the compared models show quite large differences in those LW net fluxes for the surface layer (lower-left) for the 300~K runs. This analysis for the transition from optically thin to optically thick is generally very sensitive to modeling parameters such as e.g. individual layer heights, used line lists and continua, or exact H$_2$O, temperature and pressure profiles. We read these profiles off the plots in \cite{yang2016}, which may have introduced extra differences, additional to the already large differences of the models originally compared in their study of up to $\sim$100\%. To further clarify this, we plan on doing a thorough k-distribution model inter-comparison in the future. For absorption of stellar radiation in the SW, both the radiation reaching the surface (upper-left), and the back-scattered radiation escaping again at TOA (upper-right) compare well with the other models.
%
\subsection{Cloud-Free Venus}
While the above tests were performed for Earth-like planets, Figure \ref{Venus-Temp} shows our temperature profiles for a cloud-free Venus calculated with 1D-TERRA in climate-only mode with prescribed atmospheric composition, compared to the Venus International Reference Atmosphere VIRA-1 \citep{seiff1985} and a cloud-free Venus model result from \citet{mendonca2015}. For this test we set the planetary parameters to represent Venus, e.g.~with surface gravity $g=8.87\rm\,m/s^2$, surface pressure $p_0=93\rm\,bar$, solar constant set to 1.913~$S_{\rm Earth}$, and due to the lack of cloud treatment in our model we set the surface albedo to 0.755. Note however that due to the thick atmosphere, an influence of the surface albedo on the planetary albedo is expected to be negligible \citep[see, e.g.,][]{vonParis2013,godolt2016}, which we also found in tests with a basaltic surface albedo of 0.13 (not shown). The atmospheric composition is prescribed with 96.5$\%$ of CO$_2$, 3.5$\%$ N$_2$, 20~ppm H$_2$, 20~ppm O$_2$, 100~ppb O$_3$, plus surface concentrations of 150~ppm SO$_2$, 32.5~ppm H$_2$O, 20~ppm CO, and 150~ppb HCl with profiles following \citet{haus2015} and \citet{tsang2008}. H$_2$O profiles are not calculated from saturation vapor pressures, but fixed to the initial concentration, as Venus does not have a surface ocean. Our model TOA is set to 0.01~Pa, which corresponds to $\sim$ 124~km, and our initial temperature profile starts at 735.3~K at the surface, linearly decreasing in height up to 169~K at the model TOA. The choice of initial temperature profile does not influence the outcome, but heavily influences the model run-time until full radiative-convective equilibrium is achieved. For Venus, instead of the requirement for full radiative-convective equilibrium, equilibrium was determined when OLR changes and temperature changes in each layer were sufficiently small over 10 Venus days (dOLR/d$t$~$<$~0.5~W/m$^2$ and d$T$/d$t$~$<$~0.1~K), similar to the approach by \citet{mendonca2015}. We see in Figure \ref{Venus-Temp} that our cloud-free Venus temperature profile compares well with the cloud-free model studies conducted by \citet{mendonca2015}, while still being remarkably close to VIRA measurements. The two features associated with the missing clouds, are the missing mid-atmospheric heating from cloud absorption and convection within, and the higher tropospheric temperatures due to the absorption of stellar radiation deeper down in the atmosphere. \\
Figure \ref{Venus-Cool} compares our cloud-free cooling rates to the gas contributions to atmospheric cooling in the VIRA atmosphere modeled by \citet{haus2015}. Note that \citet{haus2015} include the effect of clouds indirectly via the use of the cloudy VIRA atmosphere,  and they use absorption data from various high-temperature line lists. Several model studies \citep[see, e.g.,][]{kopparapu2013} have shown that cooling rates and OLRs obtained with HITRAN vs. HITEMP line lists start to diverge above $\sim$800-1000~K. As Venus surface temperatures approach those temperatures, ideally HITEMP or Exomol should be used. While both model rates follow similar trends, our cloud-free model outputs somewhat lower cooling rates, dominated by the two regions, around 1 bar and slightly above 0.1~bar, where temperature differences increase, mainly related to the missing cloud layers in our model. Lastly, we conduct a detailed TOA OLR analysis for our model temperature and composition profile, band by band against the LBL code GARLIC, shown in Figure \ref{Venus-valid-fluxes}. Results suggest excellent agreement for $F\rm_{up}$ fluxes in all CO$_2$ and CO bands, slightly less absorption in REDFOX for SO$_2$, and somewhat more absorption in the LW H$_2$O continuum above 20~$\mu\rm m$. There are two apparent window regions with generally low absorption in Venus' atmosphere, where REDFOX $F\rm_{up}$ fluxes show significantly more (10-20$\%$) absorption. These are the 6~$\mu\rm m$ H$_2$O band, and the 3.4~$\mu\rm m$ band, a window region with negligible contributions from absorbers. The total OLR for calculations using the Gauss-Legendre quadratures (549.64~W/m$^2$) agree within $\sim$1$\%$ with the LBL calculations (556.01~W/m$^2$). Figure \ref{Venus-valid-fluxes} also shows fluxes for REDFOX with the double Gaussian quadratures (blue) for comparison, to confirm that our choice of using the Gauss-Legendre quadratures is also justified for Venus, and Venus-like planets. \\
In conclusion for our Venus tests, results with REDFOX show slightly lower cooling rates (hence long-wave absorption) than the LBL model used by \citet{haus2015}, yet slightly higher overall long-wave absorption than the LBL model GARLIC, and temperatures compare well to measurements and other cloud-free model studies. In summary we have shown that Venus temperatures, calculated with our new general-purpose climate model, compare very well with observations, and with other model studies designed to reproduce Venus.
%
\subsection{Aerosol-Free Mars}
Our final validation, shown in Fig.~\ref{mars-temp}, is for an aerosol-free modern Mars scenario (orange). We compare temperatures modeled with 1D-TERRA in coupled climate-chemistry mode against the 
reference atmosphere of \citet{haberle2017} (black) for a low-dust scenario based on diurnal averages of observations from the MCS instrument \citep{Kleinboehl2009}. Also overplotted (green) are zonally-averaged temperature ranges based on Mariner 9 IRIS data \citep{justus1996}. As initial condition we set a 250~K isoprofile, Martian atmospheric composition of major species is taken from \citet{owen1977}, the water vapor profile and the profiles of minor species are taken from \citet{nair1994}. Mars' planetary albedo of 0.29 \citep{clawson1991} is for most of the time throughout a Martian year (i.e. without global dust storms) dominated by the surface. Our cloud-free assumption in radiative transfer using a surface albedo of 0.290 is therefore reasonable. As Mars does not have a surface water ocean reservoir, we do not calculate tropospheric H$_2$O profiles from RH profiles.\\
Our model results shown in Fig.~\ref{mars-temp} compare reasonably well with Martian lower and upper atmospheric temperatures, while middle atmospheric temperatures are 10-30~K lower compared to Mars reference data. We are not parameterizing Mie scattering from aerosols in this work, which would lead to extra heating in the Martian lower to middle atmosphere, as discussed by e.g. \citet{gierasch1972}. 
%
%
\section{Application to K2-18b} \label{applications}
%
We apply 1D-TERRA to the super-Earth / sub-Neptune planet K2-18b with recently analyzed atmospheric infrared features from combined Kepler K2, Hubble WFC3, and Spitzer observations \citep{benneke2019,Tsiaras2019}. This marks the first detection of H$_2$O in the atmosphere of an exoplanet smaller than Neptune and Uranus. K2-18b has a mass of 8.63~$\pm$~1.35~M$_\oplus$ and orbits an M2.5 host star (0.0234~L$_\odot$) at 0.1429~AU with resulting Total Stellar Irradiation (TSI)~=~1.1459~TSI$_\oplus$. For the host star, K2-18, we use the stellar spectrum from GJ176, an M2.5 star similar to K2-18 in stellar properties, available in the MUSCLES database, and prepare it as decribed in Section~\ref{sssec:sources}. K2-18b's planetary radius was recently updated from 2.27 to 2.711~$\pm$~0.065~R$_\oplus$ due to reanalysis of K2-18's stellar radius by \citet{cloutier2019}. Planetary radius calculations are based on the 4.5~$\mu\rm m$ Spitzer band transit depths. We apply a simple mass-radius parameterization from \citet{noack2016} to estimate gravity at the surface-atmosphere boundary layer and find that the original radius estimate of 2.27~R$_\oplus$ could still be interpreted as a rocky planet with an ice-mass-fraction of 35~-~62\% and a core-mass-fraction $<$~35\%. These estimates lead to a surface gravity of $g\,\sim\rm 15\,m/s^2$. The updated radius of 2.71~R$_\oplus$, however, cannot be achieved using this parameterization without a significant atmospheric contribution.\\
We compare a variety of planetary scenarios and atmospheres calculated with the extensive capabilities of 1D-TERRA against the published observations. Our main goals here are to show the large application range of our new model, and to provide possible explanations and constraints when interpreting the observed spectral features of K2-18b. 
%
\subsection{Scenarios}
\begin{deluxetable*}{l c c c c c c}
    \tabletypesize{\footnotesize}
    \tablecolumns{7}
    \tablecaption{Planetary and atmospheric scenarios for modeled K2-18b. \label{tbl:scenarios}}
    \tablehead{\colhead{run} & \colhead{$p_0$ start [bar]} & \colhead{H$_2$O reservoir} & \colhead{$p_0$ end [bar]} & \colhead{atm. composition} & \colhead{$T$ start} & \colhead{$T(p_0)$ end [K]}} 
    \startdata
        (1) Earth-like & 1.56 & Yes & 132.2 & US Std. 1976 & US Std. 1976 & 604\\
        (2) Venus-like & 143.6 & No & 143.6 & VIRA-1 & ISO 500~K & 550\\
        (3) Venus + 186~bar H$_2$ & 93+186 & No & 279 &  VIRA-1 + H$_2$ & ISO 500~K & 699\\
        (4) Venus + 800~bar H$_2$ & 93+800 & No & 893 &  VIRA-1 + H$_2$ & ISO 500~K & 893\\
        \hline
        (5) Solar Metallicity & 10 & No & 10 & CEA(1xSolar) & Linear 320 to 250~K & 691\\
        (6) Solar Met. + H$_2$O Ocean & 10 & Yes & 230.2 & CEA(1xSolar) + H$_2$O & Linear 320 to 250~K & 925\\
        (7) 50x Solar Metallicity & 10 & No & 10 &  CEA(50xSolar) & Linear 520 to 250~K & 719 \\
    \enddata
\end{deluxetable*}
\paragraph{(1)} For our first scenario shown in Tbl.~\ref{tbl:scenarios} we start with an Earth-like atmosphere based on the US standard atmosphere 1976. As a starting condition we assume Earth's atmospheric mass, which, with the elevated gravity, leads to a surface pressure of 1.55~bar. Recent modeling studies have shown that surface temperatures of Earth-like planets around M-stars would increase relative to Earth for Earth's TSI, because of the red-shifted spectrum of cooler stars, different UV environment, and atmospheric and ocean responses \citep[see e.g.][]{Scheucher_2018}. Consequently, we expect a theoretically Earth-like K2-18b, receiving somewhat higher TSI than Earth, to significantly warm up, evaporating parts of the ocean H$_2$O reservoir, resulting in a thick steam atmosphere.  For this reason we use a constant relative humidity profile of 0.95 used in Eq.~\ref{eq:h2ocalc} for the H$_2$O calculation over all convective layers. For this scenario we use 1D-TERRA in fully coupled climate-chemistry mode to capture details of the impact of increased H$_2$O in the atmosphere.
\paragraph{(2)} Due to the higher TSI and the size of the planet, the accumulation of a significant atmosphere and a past runaway greenhouse state are plausible, therefore our second atmosphere of interest is an Exo-Venus. As with Venus, we do not assume an H$_2$O reservoir and, to retain Venus' atmospheric mass, we increase the surface pressure accordingly to 143.6~bar. All Venusian scenarios (2-4) are run in climate-only mode, as we focus here on the resulting H$_2$O features in the modeled transmission spectra influenced by temperatures, pressures, and H$_2$O amounts, rather than the full photo-chemistry of other trace gases. Our initial temperatures for the Venus-like runs are 500~K iso-profiles.
\paragraph{(3-4)} To increase the amplitude of absorption features, we increase the atmospheric scale heights by adding H$_2$ to our Exo-Venus atmospheres. First we triple (=~279~bar p$_0$) our original (=~93~bar p$_0$) Venus atmosphere by adding 186~bar of H$_2$ (3), and in another scenario we add 800~bar of H$_2$ on top of the 93~bar of Venus for comparison (4). This was chosen to stay within our limit of calculated k-distributions of 1000~bar.
\paragraph{(5)} Next, we move on to hydrogen-dominated primary atmospheres, which we model with 1D-TERRA in climate-only mode with prescribed atmospheric composition. We have calculated the equilibrium chemical composition for an isothermal atmosphere with an equilibrium temperature of $T\rm{_{eq}=320\,K}$ using the NASA CEA model \citep[][]{CEA1} with initial solar elemental abundances by \citet{Asplund2009}.  The chemical composition is subsequently kept fixed. Radiative-convective equilibrium is calculated down to 10 bars, although the atmosphere might extend further down to higher pressures. We choose this limit, because we expect H$_2$O absorption features in the Hubble WFC3 spectral range to result from pressures up to a few hundred mbar and we cannot make any assumption on the actual extent down to a possible surface. For all H$_2$ dominated atmospheres we start with a linear temperature gradient from the surface up to the TOA, in order to speed up climate calculations.
\paragraph{(6)} We repeat the same with a potential H$_2$O reservoir (ocean) at the lower atmospheric boundary at 10 bars which can evaporate and mix into the atmosphere dependent on temperature, H$_2$O saturation pressure, and assumed constant relative humidity of 95\% for convective layers as in scenario \emph{(1)}, due to the expected warm temperatures in the lower atmosphere. To stay well within our valid model temperature range ($T<1000$~K) we limit the H$_2$O saturation vapor pressure, i.e. the amount of water the atmosphere can hold in gas phase, to its value at the critical point at 647~K for temperatures above, leading to a maximum of 220~bars of H$_2$O in the atmosphere. As in scenario \emph{(5)} the chemical composition is kept fixed, except for the changes in H$_2$O, which changes all mixing ratios accordingly.
\paragraph{(7)} Lastly, we calculate a 50 times solar metallicity atmosphere with the NASA CEA model and perform climate calculations similar to scenario \emph{(5)} without any H$_2$O reservoir to study the effect of metallicity on atmospheric scale heights, impacting the amplitude of resulting absorption features. \\
%
\subsection{Results} \label{sec:results}
\picwide{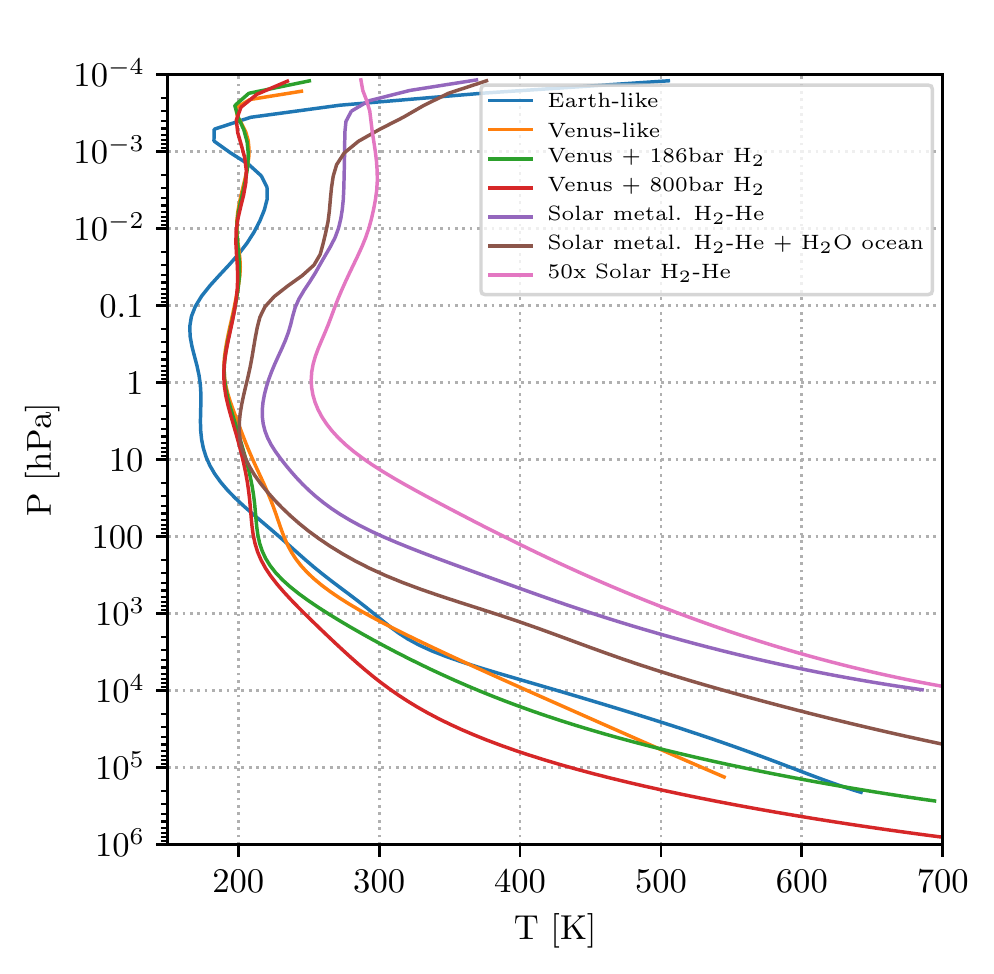}{Temperature profiles for our K2-18b model atmospheres (see Tbl.~\ref{tbl:scenarios}).}
\picfullwide{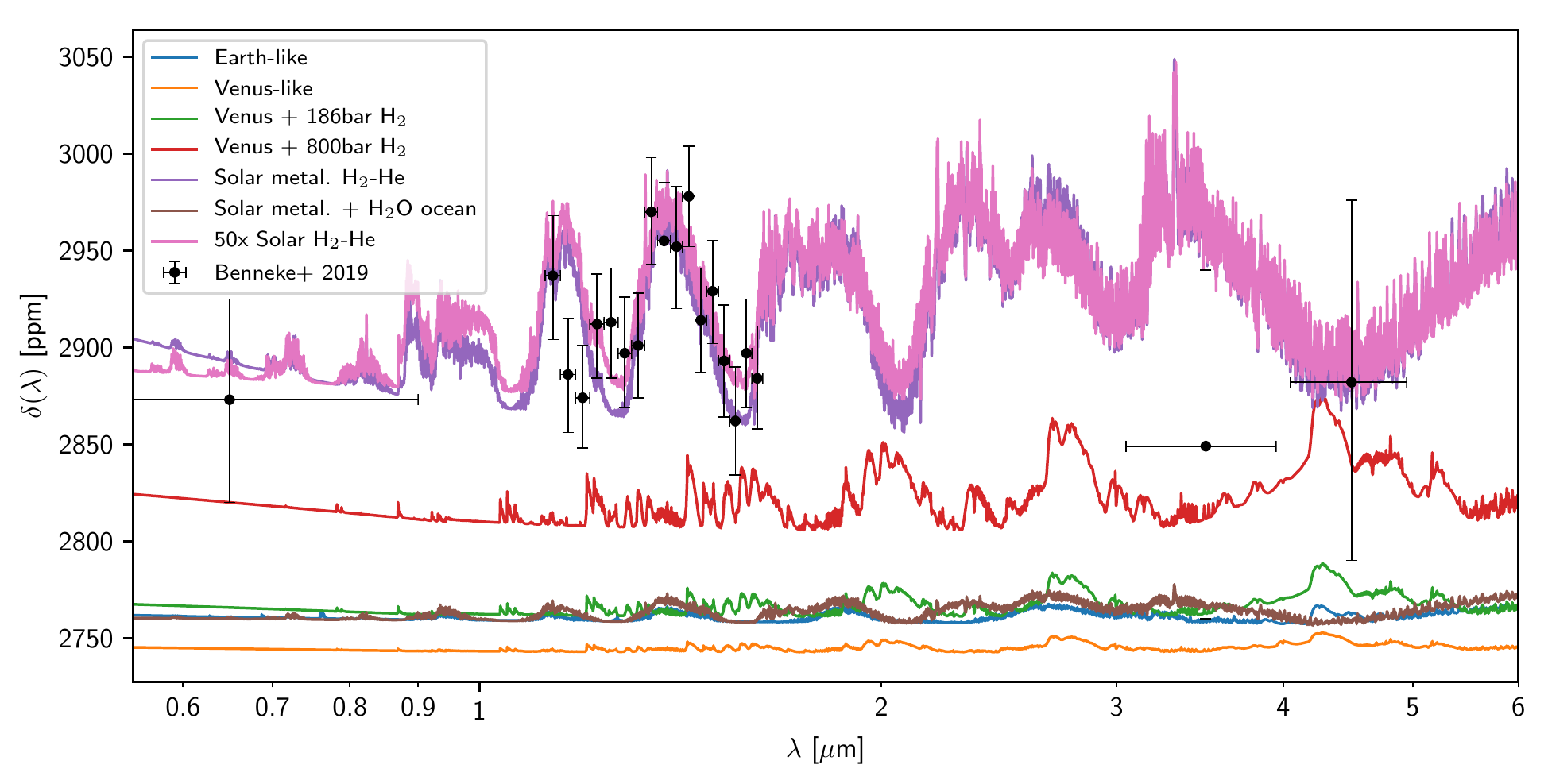}{Comparison of planetary transmission spectra for K2-18b calculated with GARLIC for our 1D-TERRA model atmospheres (Tbl.~\ref{tbl:scenarios} and Fig.~\ref{k218b-temp}) with Kepler, Hubble, and Spitzer data taken from \citet{benneke2019}. $\delta(\lambda)$ is the wavelength-dependent transit depth of the planet.}
\picwide{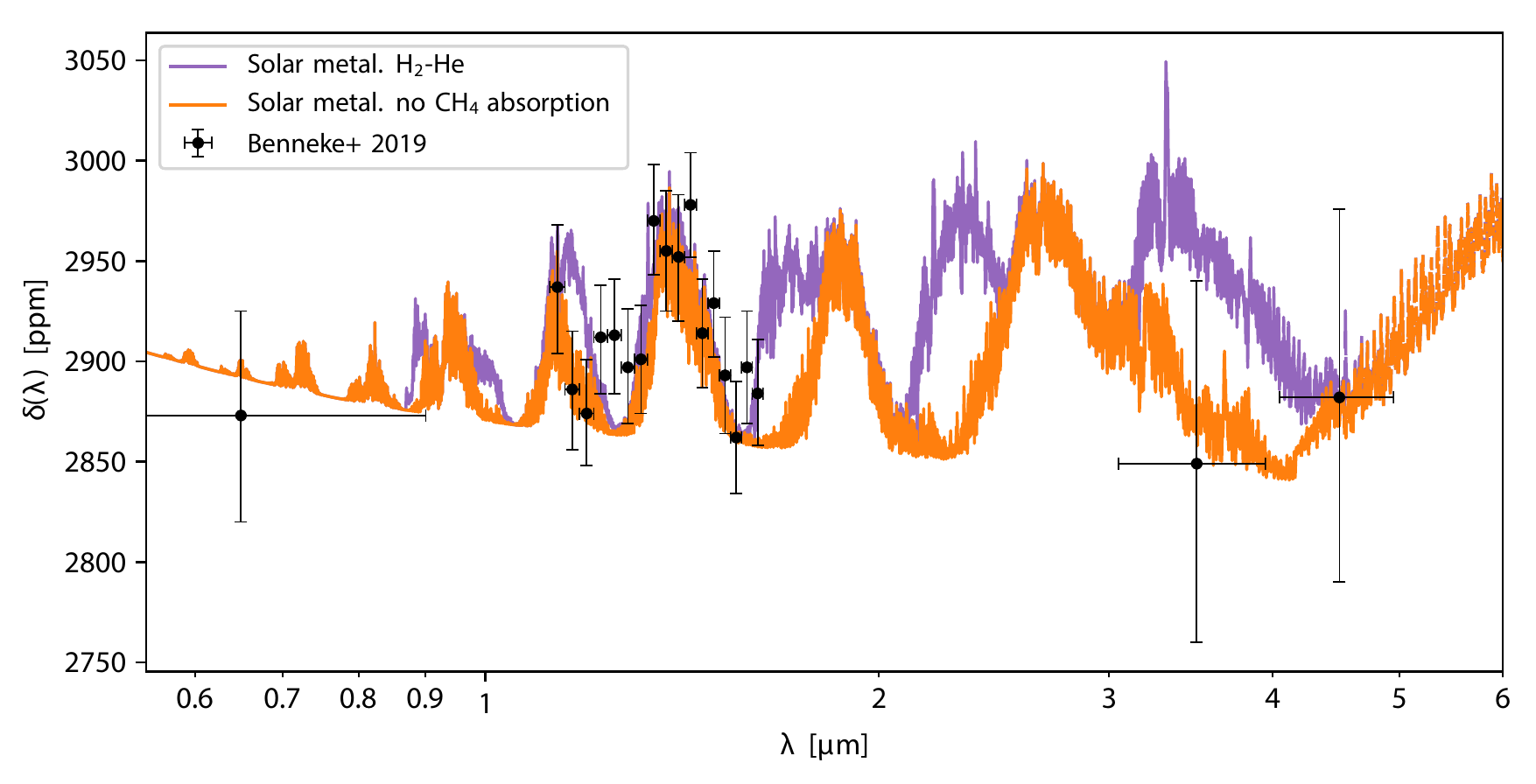}{Transmission spectra for the solar metallicity atmosphere (purple) as in Fig.~\ref{k218b-spec} compared to the same atmosphere, but without CH$_4$ contribution (orange).}
Table~\ref{tbl:scenarios} shows the different planetary scenarios modeled in this work, plus the resulting Bottom-Of-Atmosphere (BOA) pressures after the climate, or climate-chemistry calculations, respectively, as well as the respective BOA temperatures. BOA pressures can only change in our model when mass conservation is broken by additional in- or outfluxes. This is invoked e.g. for scenarios with a H$_2$O reservoir, evaporating extra molecules into the atmosphere when the atmosphere is heating up, or condensing them out.\\
Fig.~\ref{k218b-temp} shows the resulting atmospheric temperature profiles for all scenarios described above. Almost all scenarios lead to global mean temperatures in the thick lower atmosphere which are too hot to be habitable. That said, at the 1~bar level all runs, except the hydrogen-dominated cases, yield moderate temperatures with the possibility of liquid water, as also discussed by \citet{benneke2019}. At higher pressures, temperature profiles become adiabatic. At low atmospheric pressures, UV heating leads to the temperature inversions in the upper layers.\\
Our main results are shown in Fig.~\ref{k218b-spec}. We supply the modeled atmospheric compositions, and temperatures from Fig.~\ref{k218b-temp}, to calculate forward modeled transmission spectra in GARLIC, and compare them with the Kepler, Hubble, and Spitzer data of K2-18b from \citet{benneke2019}. We plot the planetary transit depth, $\delta(\lambda)=[(r_p + h(\lambda))/r_s]^2$,
with $r_p$ being the planetary radius, $r_s$ the stellar radius, and  $h(\lambda)=\sum_i(1-\mathcal{T}_i(\lambda))\Delta h_i$ the effective height of the atmosphere for a given wavelength. Note that all spectra shown can in principle be shifted vertically within the error in the 4.5~$\mu\rm m$ Spitzer observation, from which radius estimates for K2-18b were derived, but it was rather our aim to focus on the amplitudes of spectral features, rather than the closest observational fit. Such a shift could arise due to the lack of knowledge regarding e.g. the interior structure and location of the planet-atmosphere boundary. Spectral absorption features in the modeled Solar composition atmosphere (purple) and the 50 times Solar metallicity results (pink) match the observations well within $\pm$~1$\sigma$, with the exception of the 3.6~$\mu\rm m$ Spitzer band. A distinction between 1 and 50 times Solar metallicities from the observations is difficult. Other considered scenarios cannot produce features strong enough to explain the apparent 1-2~$\mu$m H$_2$O absorption observations with a strength of $\sim$100~ppm and a 1$\sigma$ of $\sim$23~ppm. For example, our H$_2$-dominated Venus-like case \emph{(4)}, shown in red, produces feature strengths of only $\sim$35~ppm.\\
Furthermore, the addition of extra H$_2$O due to a hypothetical water world ocean reservoir (brown) would increase atmospheric mean molecular weights, thus decrease scale heights, drastically. For the Solar metallicity case with ocean reservoir \emph{(6)}, features would reduce to $<$15~ppm. Even if one were to shift these spectra vertically in Fig.~\ref{k218b-spec}, it is clear that the amplitude of features seen in any model scenarios (other than the H$_2$-He atmospheres without an ocean reservoir) would still not be compatible with the observations.
A further test run with an Earth-like relative humidity profile with 80\% humidity at the surface and a very dry upper atmosphere to calculate the H$_2$O profile after \citet{manabe1967} suggested no significant changes in planetary climate and atmospheric spectral appearance. \citet{Madhusudhan2020} recently argued that the amount of H$_2$O mixed into an H$_2$/He atmosphere consistent with the HST measurements would not significantly change the planetary radius, and further that a habitable ocean world with liquid water underneath a H$_2$/He atmosphere is possible for K2-18b. While it may be true that the observed radius can be achieved in such a scenario, our study considers the effect such a large H2O reservoir would have upon the atmosphere due to strong evaporation and thereby on the observed spectral features. Our results suggest that an ocean world scenario would not fit the observed spectral features.
This includes all our terrestrial atmospheres, and also the hypothetical Venus runs with added H$_2$, although the Venus run with 800~bars of added H$_2$ (red) already shows distinguishable features. While it is clear that larger scale heights are needed to explain the observations, it is interesting that our results suggest that large amounts of the heavy molecule CO$_2$ can also be excluded by observations, as well as the existence of a hypothetical surface ocean at our BOA temperatures. \\
The contrast gradient between the two Spitzer observations at 3.6 and 4.5~$\mu\rm m$ could come from extensive amounts of CO$_2$ or other absorbers in the 4.5~$\mu\rm m$ band, e.g. O$_3$, N$_2$O, or CO (all not shown), but all of which would further bring down scale heights. Another possibility is reduced CH$_4$ in the middle and upper atmosphere, e.g. by freezing it out (at $\sim91\rm\,K$), and in doing so reducing the 3.6~$\mu\rm m$ CH$_4$ absorption feature. The CEA Solar composition atmosphere contains $\sim 460\rm\,ppm$ CH$_4$ (50x Solar $\sim$2.48$\%$ CH$_4$). In Fig. \ref{k218b-spec-noch4} we show how a transmission spectrum without CH$_4$ contribution would look by comparison. Our results suggest that the depletion of CH$_4$ could indeed explain the discrepancy between model result and observations. \citet{benneke2019} could only limit CH$_4$ to $<$~0.25\%, while our results with 1D-TERRA limit CH$_4$ to $<460\rm\,ppm$.
%
%
\section{Summary} \label{sec:sum}
%
We describe in this work our new radiative transfer module, REDFOX, implemented in our 1D climate-chemistry model, 1D-TERRA. 1D-TERRA also uses the updated photo-chemistry reaction scheme, BLACKWOLF, which is described in depth in the companion paper \citet[][in prep.]{wunderlich2019b}. The radiative transfer update included the user's choice of absorbing species, extended wavelength ranges of stellar and thermal calculations, flexible inclusion of CIAs, additional continua, and UV absorbers, and broader pressure - temperature ranges, amongst others. We show an extensive verification and validation of REDFOX and 1D-TERRA against terrestrial-type Solar System atmospheres, and previously published modeling results for other theoretical atmospheres. \\ 
With our new model we study the exoplanet K2-18b, with its first ever observed atmospheric absorption features of H$_2$O in an exoplanet with mass $<$~10~M$_\oplus$, and calculate climate and transmission spectra for different atmospheric and planetary scenarios.\\
We confirm the findings of \citet{benneke2019} that K2-18b likely has a H$_2$-He atmosphere, and additionally exclude the existence of a large H$_2$O ocean reservoir exposed to the atmosphere. This rules out a water layer at the top of the mantle for possible interior structures, as it would decrease the atmospheric scale height, despite the newly described effect of increased radii for H$_2$O steam atmospheres by \citet{turbet2019}. We further confirm that the Hubble WFC3 observations between 1 - 2~$\mu\rm m$ can be explained by H$_2$O absorption in a hydrogen-dominated primary atmosphere. Both Solar and 50 times Solar metallicity results can explain the observations alike. With the recent radius update for K2-18b, and the need for a significant atmospheric contribution to the planetary radius, a hypothetical H$_2$O reservoir would likely be a super-critical fluid at such high temperatures implying an envelope rather than a solid, or liquid surface. This suggests that K2-18b is unlikely to be a water world or even an icy world. We also argue that significant amounts of CH$_4$ are excluded by the Spitzer observations. The depletion of CH$_4$ has also been discussed in depth for the significantly larger Neptune-sized exoplanet GJ436b \citep[see e.g.][]{Madhusudhan2011,Hu2015} and recently by \citet{benneke2019gj3470b} for the 12~M$_\oplus$ sub-Neptune GJ3470b, while Neptune itself shows significant amounts of CH$_4$ with around 1.5\% in the atmosphere.\\
Additionally our results clearly show in comparison how small atmospheric contributions would be if K2-18b were to possess a terrestrial atmosphere, such as Earth-like or Venus-like. Even with large added amounts of H$_2$ to increase scale heights, spectral features are nevertheless suppressed within the error-bars of state-of-the-art transit observations, which suggests that it will be challenging for next-generation telescopes to characterize Earth- or Venus-like atmospheres from primary transits. That said, we are excited at the prospect of JWST characterizing many more super-Earth atmospheres in the near future.\\
%
\acknowledgments
We thank Nadine Nettelmann for her constructive inputs and fruitful discussions and the anonymous referee for his/her constructive comments that helped improve and clarify the manuscript\\
M.~Scheucher acknowledges support from DFG project RA 714/9-1. \\
F.~Wunderlich acknowledges support from DFG project RA 714/7-1 \\
M.~Godolt gratefully acknowledges support from the DFG through the project GO 2610/1-1 and the priority programme SPP 1992 ``Exploring the Diversity of Extrasolar Planets'' (GO 2610/2-1). \\
K.~Herbst and J.~Lee~Grenfell acknowledge the International Space Science Institute and the supported International Team 464: \textit{The  Role  Of  Solar  And  Stellar  Energetic  Particles  On (Exo)Planetary Habitability (ETERNAL, \url{http://www.issibern.ch/teams/exoeternal/})}.\\
F.~Schreier acknowledges support from DFG project SCHR 1125/3-1.\\

\software{GARLIC \citep{schreier2014,schreier2018agk,schreier2018ace}, HITRAN2016 \citep{GORDON20173}}, MPI Mainz Spectral Atlas \citep{spectralatlas}, KSPECTRUM1.2.0 \citep{Eymet_2016}, 1D Climate-Chemistry model legacy \citep[][ and others]{Kasting1986,pavlov2000,segura2003,vonparis2015}, NASA CEA \citep{CEA1}

\bibliographystyle{aasjournal}

\begin{thebibliography}{}
\expandafter\ifx\csname natexlab\endcsname\relax\def\natexlab#1{#1}\fi
\providecommand{\url}[1]{\href{#1}{#1}}

\bibitem[{Abel {et~al.}(2011)Abel, Frommhold, Li, \& Hunt}]{abel2011}
Abel, M., Frommhold, L., Li, X., \& Hunt, K. L.~C. 2011, The Journal of
  Physical Chemistry A, 115, 6805, pMID: 21207941.
\newblock \url{https://doi.org/10.1021/jp109441f}

\bibitem[{Abel {et~al.}(2012)Abel, Frommhold, Li, \& Hunt}]{abel2012}
---. 2012, The Journal of Chemical Physics, 136, 044319.
\newblock \url{https://doi.org/10.1063/1.3676405}

\bibitem[{{Abramowitz} \& {Stegun}(1972)}]{abramowitz1972}
{Abramowitz}, M., \& {Stegun}, I.~A. 1972, {Handbook of Mathematical Functions}

\bibitem[{{Allen}(1973)}]{allen1973}
{Allen}, C.~W. 1973, {Astrophysical quantities}

\bibitem[{Amundsen {et~al.}(2017)Amundsen, Tremblin, Manners, Baraffe, \&
  Mayne}]{amundsen2016}
Amundsen, D.~S., Tremblin, P., Manners, J., Baraffe, I., \& Mayne, N.~J. 2017,
  A\&A, 598, A97.
\newblock \url{https://doi.org/10.1051/0004-6361/201629322}

\bibitem[{Anglada-Escud{\'e} {et~al.}(2016)Anglada-Escud{\'e}, Amado, Barnes,
  Berdi{\~n}as, Butler, Coleman, De~La~Cueva, Dreizler, Endl, Giesers,
  {et~al.}}]{anglada2016}
Anglada-Escud{\'e}, G., Amado, P.~J., Barnes, J., {et~al.} 2016, Nature, 536,
  437

\bibitem[{{Asplund} {et~al.}(2009){Asplund}, {Grevesse}, {Sauval}, \&
  {Scott}}]{Asplund2009}
{Asplund}, M., {Grevesse}, N., {Sauval}, A.~J., \& {Scott}, P. 2009, \araa, 47,
  481

\bibitem[{Baranov {et~al.}(2003)Baranov, Fraser, Lafferty, \&
  Vigasin}]{baranov2003}
Baranov, Y.~I., Fraser, G.~T., Lafferty, W.~J., \& Vigasin, A.~A. 2003, in
  Weakly Interacting Molecular Pairs: Unconventional Absorbers of Radiation in
  the Atmosphere, ed. C.~Camy-Peyret \& A.~A. Vigasin (Dordrecht: Springer
  Netherlands), 149--158

\bibitem[{Baranov {et~al.}(2004)Baranov, Lafferty, \& Fraser}]{BARANOV2004432}
Baranov, Y.~I., Lafferty, W., \& Fraser, G. 2004, Journal of Molecular
  Spectroscopy, 228, 432 , special Issue Dedicated to Dr. Jon T. Hougen on the
  Occasion of His 68th Birthday.
\newblock
  \url{http://www.sciencedirect.com/science/article/pii/S0022285204001390}

\bibitem[{Benneke {et~al.}(2019{\natexlab{a}})Benneke, Wong, Piaulet, Knutson,
  Lothringer, Morley, Crossfield, Gao, Greene, Dressing, Dragomir, Howard,
  McCullough, Kempton, Fortney, \& Fraine}]{benneke2019}
Benneke, B., Wong, I., Piaulet, C., {et~al.} 2019{\natexlab{a}}, The
  Astrophysical Journal Letters, 887, L14.
\newblock \url{https://doi.org/10.3847%2F2041-8213%2Fab59dc}

\bibitem[{Benneke {et~al.}(2019{\natexlab{b}})Benneke, Knutson, Lothringer,
  Crossfield, Moses, Morley, Kreidberg, Fulton, Dragomir, Howard,
  {et~al.}}]{benneke2019gj3470b}
Benneke, B., Knutson, H.~A., Lothringer, J., {et~al.} 2019{\natexlab{b}},
  Nature Astronomy, 3, 813

\bibitem[{Boone {et~al.}(2005)Boone, Nassar, Walker, Rochon, McLeod, Rinsland,
  \& Bernath}]{boone2005}
Boone, C.~D., Nassar, R., Walker, K.~A., {et~al.} 2005, Appl. Opt., 44, 7218.
\newblock \url{http://ao.osa.org/abstract.cfm?URI=ao-44-33-7218}

\bibitem[{Bucholtz(1995)}]{bucholtz1995}
Bucholtz, A. 1995, Appl. Opt., 34, 2765.
\newblock \url{http://ao.osa.org/abstract.cfm?URI=ao-34-15-2765}

\bibitem[{Burkholder {et~al.}(2015)Burkholder, Sander, Abbatt, Barker, Huie,
  Kolb, Kurylo, Orkin, Wilmouth, \& Wine}]{burkholder2015}
Burkholder, J., Sander, S., Abbatt, J., {et~al.} 2015, Chemical kinetics and
  photochemical data for use in atmospheric studies: evaluation number 18,
  Tech. rep., Pasadena, CA: Jet Propulsion Laboratory, National Aeronautics and
  Space~…

\bibitem[{Charbonneau {et~al.}(2009)Charbonneau, Berta, Irwin, Burke, Nutzman,
  Buchhave, Lovis, Bonfils, Latham, Udry, {et~al.}}]{charbonneau2009}
Charbonneau, D., Berta, Z.~K., Irwin, J., {et~al.} 2009, Nature, 462, 891

\bibitem[{Chase(1998)}]{chase1998}
Chase, M.~W. 1998, J. Phys. Chem. Ref. Data Monographs and Supplements, 9, 1

\bibitem[{Clawson \& (U.S.)(1991)}]{clawson1991}
Clawson, J.~F., \& (U.S.), J. P.~L. 1991, Thermal environments No. JPL D-8160
  ([Pasadena, Calif.]: [Jet Propulsion Laboratory, California Institute of
  Technology])

\bibitem[{Clough {et~al.}(1992)Clough, Iacono, \& Moncet}]{clough1992}
Clough, S.~A., Iacono, M.~J., \& Moncet, J.-L. 1992, Journal of Geophysical
  Research: Atmospheres, 97, 15761.
\newblock
  \url{https://agupubs.onlinelibrary.wiley.com/doi/abs/10.1029/92JD01419}

\bibitem[{{Clough} {et~al.}(2005){Clough}, {Shephard}, {Mlawer}, {Delamere},
  {Iacono}, {Cady-Pereira}, {Boukabara}, \& {Brown}}]{clough2005}
{Clough}, S.~A., {Shephard}, M.~W., {Mlawer}, E.~J., {et~al.} 2005, \jqsrt, 91,
  233

\bibitem[{Cloutier {et~al.}(2019)Cloutier, Astudillo-Defru, Doyon, Bonfils,
  Almenara, Bouchy, Delfosse, Forveille, Lovis, Mayor, Menou, Murgas, Pepe,
  Santos, Udry, \& W\"unsche}]{cloutier2019}
Cloutier, R., Astudillo-Defru, N., Doyon, R., {et~al.} 2019, A\&A, 621, A49.
\newblock \url{https://doi.org/10.1051/0004-6361/201833995}

\bibitem[{Cronin(2014)}]{cronin2014}
Cronin, T.~W. 2014, Journal of the Atmospheric Sciences, 71, 2994.
\newblock \url{https://doi.org/10.1175/JAS-D-13-0392.1}

\bibitem[{Crossfield \& Kreidberg(2017)}]{crossfield2017}
Crossfield, I.~J., \& Kreidberg, L. 2017, arXiv preprint arXiv:1708.00016

\bibitem[{Deming \& Shupe(1931)}]{deming1931}
Deming, W.~E., \& Shupe, L.~E. 1931, Phys. Rev., 38, 2245.
\newblock \url{https://link.aps.org/doi/10.1103/PhysRev.38.2245}

\bibitem[{Edl{\'{e}}n(1966)}]{edlen1966}
Edl{\'{e}}n, B. 1966, Metrologia, 2, 71.
\newblock \url{https://doi.org/10.1088%2F0026-1394%2F2%2F2%2F002}

\bibitem[{Elsasser(1942)}]{elsasser1942}
Elsasser, W. 1942, Heat Transfer by Infrared Radiation in the Atmosphere,
  Harvard meteorological studies (Harvard University, Blue Hill meteorological
  observatory).
\newblock \url{https://books.google.de/books?id=RtYgAAAAMAAJ}

\bibitem[{Eymet {et~al.}(2016)Eymet, Coustet, \& Piaud}]{Eymet_2016}
Eymet, V., Coustet, C., \& Piaud, B. 2016, Journal of Physics: Conference
  Series, 676, 012005.
\newblock \url{https://doi.org/10.1088%2F1742-6596%2F676%2F1%2F012005}

\bibitem[{Forget \& Leconte(2014)}]{forget2014}
Forget, F., \& Leconte, J. 2014, Philosophical Transactions of the Royal
  Society A: Mathematical, Physical and Engineering Sciences, 372, 20130084

\bibitem[{Gamache {et~al.}(2017)Gamache, Roller, Lopes, Gordon, Rothman,
  Polyansky, Zobov, Kyuberis, Tennyson, Yurchenko, Császár, Furtenbacher,
  Huang, Schwenke, Lee, Drouin, Tashkun, Perevalov, \&
  Kochanov}]{GAMACHE201770}
Gamache, R.~R., Roller, C., Lopes, E., {et~al.} 2017, Journal of Quantitative
  Spectroscopy and Radiative Transfer, 203, 70 , hITRAN2016 Special Issue.
\newblock
  \url{http://www.sciencedirect.com/science/article/pii/S0022407317301516}

\bibitem[{Gebauer {et~al.}(2018)Gebauer, Grenfell, Lehmann, \&
  Rauer}]{gebauer2018}
Gebauer, S., Grenfell, J., Lehmann, R., \& Rauer, H. 2018, Astrobiology, 18,
  856, pMID: 30035637.
\newblock \url{https://doi.org/10.1089/ast.2017.1723}

\bibitem[{Gebauer {et~al.}(2017)Gebauer, Grenfell, Stock, Lehmann, Godolt, von
  Paris, \& Rauer}]{gebauer2017}
Gebauer, S., Grenfell, J., Stock, J., {et~al.} 2017, Astrobiology, 17, 27,
  pMID: 28103105.
\newblock \url{https://doi.org/10.1089/ast.2015.1384}

\bibitem[{Gierasch \& Goody(1972)}]{gierasch1972}
Gierasch, P.~J., \& Goody, R.~M. 1972, Journal of the Atmospheric Sciences, 29,
  400

\bibitem[{Gillon {et~al.}(2017)Gillon, Triaud, Demory, Jehin, Agol, Deck,
  Lederer, De~Wit, Burdanov, Ingalls, {et~al.}}]{gillon2017}
Gillon, M., Triaud, A.~H., Demory, B.-O., {et~al.} 2017, Nature, 542, 456

\bibitem[{Godolt {et~al.}(2016)Godolt, Grenfell, Kitzmann, Kunze, Langematz,
  Patzer, Rauer, \& Stracke}]{godolt2016}
Godolt, M., Grenfell, J.~L., Kitzmann, D., {et~al.} 2016, \aap, 592, A36

\bibitem[{Godolt {et~al.}(2019)Godolt, Tosi, Stracke, Grenfell, Ruedas, Spohn,
  \& Rauer}]{godolt2019}
Godolt, M., Tosi, N., Stracke, B., {et~al.} 2019, A\&A, 625, A12.
\newblock \url{https://doi.org/10.1051/0004-6361/201834658}

\bibitem[{Gordon {et~al.}(2017)Gordon, Rothman, Hill, Kochanov, Tan, Bernath,
  Birk, Boudon, Campargue, Chance, Drouin, Flaud, Gamache, Hodges, Jacquemart,
  Perevalov, Perrin, Shine, Smith, Tennyson, Toon, Tran, Tyuterev, Barbe,
  Császár, Devi, Furtenbacher, Harrison, Hartmann, Jolly, Johnson, Karman,
  Kleiner, Kyuberis, Loos, Lyulin, Massie, Mikhailenko, Moazzen-Ahmadi,
  Müller, Naumenko, Nikitin, Polyansky, Rey, Rotger, Sharpe, Sung, Starikova,
  Tashkun, Auwera, Wagner, Wilzewski, Wcisło, Yu, \& Zak}]{GORDON20173}
Gordon, I., Rothman, L., Hill, C., {et~al.} 2017, Journal of Quantitative
  Spectroscopy and Radiative Transfer, 203, 3 , hITRAN2016 Special Issue.
\newblock
  \url{http://www.sciencedirect.com/science/article/pii/S0022407317301073}

\bibitem[{Grenfell {et~al.}(2007)Grenfell, Grießmeier, Patzer, Rauer, Segura,
  Stadelmann, Stracke, Titz, \& Von~Paris}]{grenfell2007}
Grenfell, J.~L., Grießmeier, J.-M., Patzer, B., {et~al.} 2007, Astrobiology,
  7, 208, pMID: 17407408.
\newblock \url{https://doi.org/10.1089/ast.2006.0129}

\bibitem[{{Grenfell} {et~al.}(2012){Grenfell}, {Grie{\ss}meier}, {von Paris},
  {Patzer}, {Lammer}, {Stracke}, {Gebauer}, {Schreier}, \&
  {Rauer}}]{grenfell2012}
{Grenfell}, J.~L., {Grie{\ss}meier}, J.-M., {von Paris}, P., {et~al.} 2012,
  Astrobiology, 12, 1109

\bibitem[{Grimm \& Heng(2015)}]{grimm2015}
Grimm, S.~L., \& Heng, K. 2015, The Astrophysical Journal, 808, 182.
\newblock \url{https://doi.org/10.1088%2F0004-637x%2F808%2F2%2F182}

\bibitem[{Gruszka \& Borysow(1997)}]{GRUSZKA1997172}
Gruszka, M., \& Borysow, A. 1997, Icarus, 129, 172 .
\newblock
  \url{http://www.sciencedirect.com/science/article/pii/S0019103597957730}

\bibitem[{Gueymard(2004)}]{gueymard2004}
Gueymard, C.~A. 2004, Solar energy, 76, 423

\bibitem[{{Gurvich} {et~al.}(1989){Gurvich}, {Veits}, \&
  {Alcock}}]{gurvich1989}
{Gurvich}, L.~V., {Veits}, I.~V., \& {Alcock}, C.~B. 1989, {Thermodynamics
  properties of individual substances. Volume 1 - Elements O, H/D, T/, F, Cl,
  Br, I, He, Ne, Ar, Kr, Xe, Rn, S, N, P, and their compounds. Part 1 - Methods
  and computation. Part 2 - Tables (4th revised and enlarged edition)}

\bibitem[{Haberle {et~al.}(2017)Haberle, Clancy, Forget, Smith, \&
  Zurek}]{haberle2017}
Haberle, R.~M., Clancy, R.~T., Forget, F., Smith, M.~D., \& Zurek, R.~W. 2017,
  The atmosphere and climate of Mars (Cambridge University Press)

\bibitem[{Hamre {et~al.}(2013)Hamre, Stamnes, Stamnes, \& Stamnes}]{hamre2013}
Hamre, B., Stamnes, S., Stamnes, K., \& Stamnes, J.~J. 2013, AIP Conference
  Proceedings, 1531, 923.
\newblock \url{https://aip.scitation.org/doi/abs/10.1063/1.4804922}

\bibitem[{Haus {et~al.}(2015)Haus, Kappel, \& Arnold}]{haus2015}
Haus, R., Kappel, D., \& Arnold, G. 2015, Planetary and Space Science, 117, 262
  .
\newblock
  \url{http://www.sciencedirect.com/science/article/pii/S0032063315002020}

\bibitem[{Hauschildt {et~al.}(1999)Hauschildt, Allard, \&
  Baron}]{hauschildt1999}
Hauschildt, P.~H., Allard, F., \& Baron, E. 1999, ApJ, 512, 377.
\newblock \url{http://stacks.iop.org/0004-637X/512/i=1/a=377}

\bibitem[{Hedges \& Madhusudhan(2016)}]{hedges2016}
Hedges, C., \& Madhusudhan, N. 2016, Monthly Notices of the Royal Astronomical
  Society, 458, 1427.
\newblock \url{https://doi.org/10.1093/mnras/stw278}

\bibitem[{Heng {et~al.}(2018)Heng, Malik, \& Kitzmann}]{heng2018}
Heng, K., Malik, M., \& Kitzmann, D. 2018, The Astrophysical Journal Supplement
  Series, 237, 29.
\newblock \url{https://doi.org/10.3847%2F1538-4365%2Faad199}

\bibitem[{Herbst {et~al.}(2019)Herbst, Grenfell, Sinnhuber, Rauer, Heber,
  Banjac, Scheucher, Schmidt, Gebauer, Lehmann, \& Schreier}]{herbst2019}
Herbst, K., Grenfell, J.~L., Sinnhuber, M., {et~al.} 2019, A\&A, 631, A101.
\newblock \url{https://doi.org/10.1051/0004-6361/201935888}

\bibitem[{Hu {et~al.}(2012)Hu, Seager, \& Bains}]{Hu2012}
Hu, R., Seager, S., \& Bains, W. 2012, The Astrophysical Journal, 761, 166.
\newblock \url{https://doi.org/10.1088%2F0004-637x%2F761%2F2%2F166}

\bibitem[{Hu {et~al.}(2015)Hu, Seager, \& Yung}]{Hu2015}
Hu, R., Seager, S., \& Yung, Y.~L. 2015, The Astrophysical Journal, 807, 8.
\newblock \url{https://doi.org/10.1088%2F0004-637x%2F807%2F1%2F8}

\bibitem[{Justus {et~al.}(1996)Justus, James, \& Johnson}]{justus1996}
Justus, C.~G., James, B.~F., \& Johnson, D.~L. 1996

\bibitem[{Karman {et~al.}(2019)Karman, Gordon, van~der Avoird, Baranov, Boulet,
  Drouin, Groenenboom, Gustafsson, Hartmann, Kurucz, Rothman, Sun, Sung,
  Thalman, Tran, Wishnow, Wordsworth, Vigasin, Volkamer, \& van~der
  Zande}]{KARMAN2019160}
Karman, T., Gordon, I.~E., van~der Avoird, A., {et~al.} 2019, Icarus, 328, 160
  .
\newblock
  \url{http://www.sciencedirect.com/science/article/pii/S0019103518306997}

\bibitem[{Kasting(1991)}]{kasting1991}
Kasting, J.~F. 1991, Icarus, 94, 1 .
\newblock
  \url{http://www.sciencedirect.com/science/article/pii/001910359190137I}

\bibitem[{{Kasting} \& {Ackerman}(1986)}]{Kasting1986}
{Kasting}, J.~F., \& {Ackerman}, T.~P. 1986, Science, 234, 1383

\bibitem[{Kasting {et~al.}(1993)Kasting, Whitmire, \& Reynolds}]{kasting1993}
Kasting, J.~F., Whitmire, D.~P., \& Reynolds, R.~T. 1993, Icarus, 101, 108

\bibitem[{Katyal {et~al.}(2019)Katyal, Nikolaou, Godolt, Grenfell, Tosi,
  Schreier, \& Rauer}]{Katyal2019}
Katyal, N., Nikolaou, A., Godolt, M., {et~al.} 2019, The Astrophysical Journal,
  875, 31.
\newblock \url{https://doi.org/10.3847%2F1538-4357%2Fab0d85}

\bibitem[{Keles {et~al.}(2018)Keles, Grenfell, Godolt, Stracke, \&
  Rauer}]{keles2018}
Keles, E., Grenfell, J.~L., Godolt, M., Stracke, B., \& Rauer, H. 2018,
  Astrobiology, 18, 116, pMID: 29364704.
\newblock \url{https://doi.org/10.1089/ast.2016.1632}

\bibitem[{Keller-Rudek {et~al.}(2013)Keller-Rudek, Moortgat, Sander, \&
  S\"orensen}]{spectralatlas}
Keller-Rudek, H., Moortgat, G.~K., Sander, R., \& S\"orensen, R. 2013, Earth
  System Science Data, 5, 365.
\newblock \url{https://www.earth-syst-sci-data.net/5/365/2013/}

\bibitem[{Kitzmann(2017)}]{kitzmann2017}
Kitzmann, D. 2017, Astronomy \& Astrophysics, 600, A111

\bibitem[{Kitzmann {et~al.}(2010)Kitzmann, Patzer, von Paris, Godolt, Stracke,
  Gebauer, Grenfell, \& Rauer}]{kitzmann2010}
Kitzmann, D., Patzer, A., von Paris, P., {et~al.} 2010, Astronomy \&
  Astrophysics, 511, A66

\bibitem[{Kleinb\"{o}hl {et~al.}(2009)Kleinb\"{o}hl, Schofield, Kass, Abdou,
  Backus, Sen, Shirley, Lawson, Richardson, Taylor, Teanby, \&
  McCleese}]{Kleinboehl2009}
Kleinb\"{o}hl, A., Schofield, J.~T., Kass, D.~M., {et~al.} 2009, Journal of
  Geophysical Research: Planets, 114, doi:10.1029/2009JE003358.
\newblock
  \url{https://agupubs.onlinelibrary.wiley.com/doi/abs/10.1029/2009JE003358}

\bibitem[{{Kopparapu} {et~al.}(2013){Kopparapu}, {Ramirez}, {Kasting}, {Eymet},
  {Robinson}, {Mahadevan}, {Terrien}, {Domagal-Goldman}, {Meadows}, \&
  {Deshpande}}]{kopparapu2013}
{Kopparapu}, R.~K., {Ramirez}, R., {Kasting}, J.~F., {et~al.} 2013, \apj, 765,
  131

\bibitem[{Lacis \& Oinas(1991)}]{lacis1991}
Lacis, A.~A., \& Oinas, V. 1991, Journal of Geophysical Research: Atmospheres,
  96, 9027.
\newblock
  \url{https://agupubs.onlinelibrary.wiley.com/doi/abs/10.1029/90JD01945}

\bibitem[{Lincowski {et~al.}(2018)Lincowski, Meadows, Crisp, Robinson, Luger,
  Lustig-Yaeger, \& Arney}]{Lincowski_2018}
Lincowski, A.~P., Meadows, V.~S., Crisp, D., {et~al.} 2018, The Astrophysical
  Journal, 867, 76.
\newblock \url{https://doi.org/10.3847%2F1538-4357%2Faae36a}

\bibitem[{Madhusudhan {et~al.}(2016)Madhusudhan, Ag{\'u}ndez, Moses, \&
  Hu}]{madhusudhan2016}
Madhusudhan, N., Ag{\'u}ndez, M., Moses, J.~I., \& Hu, Y. 2016, Space science
  reviews, 205, 285

\bibitem[{Madhusudhan {et~al.}(2020)Madhusudhan, Nixon, Welbanks, Piette, \&
  Booth}]{Madhusudhan2020}
Madhusudhan, N., Nixon, M.~C., Welbanks, L., Piette, A. A.~A., \& Booth, R.~A.
  2020, The Astrophysical Journal, 891, L7.
\newblock \url{https://doi.org/10.3847%2F2041-8213%2Fab7229}

\bibitem[{Madhusudhan \& Seager(2011)}]{Madhusudhan2011}
Madhusudhan, N., \& Seager, S. 2011, The Astrophysical Journal, 729, 41.
\newblock \url{https://doi.org/10.1088%2F0004-637x%2F729%2F1%2F41}

\bibitem[{Malik {et~al.}(2017)Malik, Grosheintz, Mendon{\c{c}}a, Grimm, Lavie,
  Kitzmann, Tsai, Burrows, Kreidberg, Bedell, Bean, Stevenson, \&
  Heng}]{Malik_2017}
Malik, M., Grosheintz, L., Mendon{\c{c}}a, J.~M., {et~al.} 2017, The
  Astronomical Journal, 153, 56.
\newblock \url{https://doi.org/10.3847%2F1538-3881%2F153%2F2%2F56}

\bibitem[{{Manabe} \& {Wetherald}(1967)}]{manabe1967}
{Manabe}, S., \& {Wetherald}, R.~T. 1967, Journal of Atmospheric Sciences, 24,
  241

\bibitem[{{McBride} \& {Gordon}(1992)}]{CEA1}
{McBride}, B., \& {Gordon}, S. 1992, Computer Program for Calculating and
  Fitting Thermodynamic Functions, Tech. rep., NASA RP-1271

\bibitem[{Meador \& Weaver(1980)}]{meador1980}
Meador, W.~E., \& Weaver, W.~R. 1980, Journal of the Atmospheric Sciences, 37,
  630.
\newblock \url{https://doi.org/10.1175/1520-0469(1980)037<0630:TSATRT>2.0.CO;2}

\bibitem[{Meadows \& Crisp(1996)}]{meadows1996}
Meadows, V.~S., \& Crisp, D. 1996, Journal of Geophysical Research: Planets,
  101, 4595.
\newblock
  \url{https://agupubs.onlinelibrary.wiley.com/doi/abs/10.1029/95JE03567}

\bibitem[{Meadows {et~al.}(2018)Meadows, Arney, Schwieterman, Lustig-Yaeger,
  Lincowski, Robinson, Domagal-Goldman, Deitrick, Barnes, Fleming, Luger,
  Driscoll, Quinn, \& Crisp}]{meadows2018}
Meadows, V.~S., Arney, G.~N., Schwieterman, E.~W., {et~al.} 2018, Astrobiology,
  18, 133, pMID: 29431479.
\newblock \url{https://doi.org/10.1089/ast.2016.1589}

\bibitem[{Mendon\c{c}a {et~al.}(2015)Mendon\c{c}a, Read, Wilson, \&
  Lee}]{mendonca2015}
Mendon\c{c}a, J., Read, P., Wilson, C., \& Lee, C. 2015, Planetary and Space
  Science, 105, 80 .
\newblock
  \url{http://www.sciencedirect.com/science/article/pii/S0032063314003444}

\bibitem[{Ment {et~al.}(2019)Ment, Dittmann, Astudillo-Defru, Charbonneau,
  Irwin, Bonfils, Murgas, Almenara, Forveille, Agol, {et~al.}}]{ment2019}
Ment, K., Dittmann, J.~A., Astudillo-Defru, N., {et~al.} 2019, The Astronomical
  Journal, 157, 32

\bibitem[{Mlawer {et~al.}(2012)Mlawer, Payne, Moncet, Delamere, Alvarado, \&
  Tobin}]{mlawer2012}
Mlawer, E.~J., Payne, V.~H., Moncet, J.-L., {et~al.} 2012, Philosophical
  Transactions of the Royal Society A: Mathematical, Physical and Engineering
  Sciences, 370, 2520.
\newblock
  \url{https://royalsocietypublishing.org/doi/abs/10.1098/rsta.2011.0295}

\bibitem[{Mlawer {et~al.}(1997)Mlawer, Taubman, Brown, Iacono, \&
  Clough}]{mlawer1997}
Mlawer, E.~J., Taubman, S.~J., Brown, P.~D., Iacono, M.~J., \& Clough, S.~A.
  1997, Journal of Geophysical Research: Atmospheres, 102, 16663.
\newblock
  \url{https://agupubs.onlinelibrary.wiley.com/doi/abs/10.1029/97JD00237}

\bibitem[{Montet {et~al.}(2015)Montet, Morton, Foreman-Mackey, Johnson, Hogg,
  Bowler, Latham, Bieryla, \& Mann}]{montet2015}
Montet, B.~T., Morton, T.~D., Foreman-Mackey, D., {et~al.} 2015, The
  Astrophysical Journal, 809, 25

\bibitem[{Murphy(1977)}]{murphy1977}
Murphy, W.~F. 1977, The Journal of Chemical Physics, 67, 5877.
\newblock \url{https://doi.org/10.1063/1.434794}

\bibitem[{Nair {et~al.}(1994)Nair, Allen, Anbar, Yung, \& Clancy}]{nair1994}
Nair, H., Allen, M., Anbar, A.~D., Yung, Y.~L., \& Clancy, R.~T. 1994, Icarus,
  111, 124

\bibitem[{Nakajima {et~al.}(1992)Nakajima, Hayashi, \& Abe}]{nakajima1992}
Nakajima, S., Hayashi, Y.-Y., \& Abe, Y. 1992, Journal of the Atmospheric
  Sciences, 49, 2256.
\newblock \url{https://doi.org/10.1175/1520-0469(1992)049<2256:ASOTGE>2.0.CO;2}

\bibitem[{Noack {et~al.}(2016)Noack, Höning, Rivoldini, Heistracher, Zimov,
  Journaux, Lammer, Hoolst, \& Bredehöft}]{noack2016}
Noack, L., Höning, D., Rivoldini, A., {et~al.} 2016, Icarus, 277, 215 .
\newblock
  \url{http://www.sciencedirect.com/science/article/pii/S001910351630149X}

\bibitem[{Owen {et~al.}(1977)Owen, Biemann, Rushneck, Biller, Howarth, \&
  Lafleur}]{owen1977}
Owen, T., Biemann, K., Rushneck, D., {et~al.} 1977, Journal of Geophysical
  research, 82, 4635

\bibitem[{Parks \& Shomate(1940)}]{parks1940}
Parks, G.~S., \& Shomate, C.~H. 1940, The Journal of Chemical Physics, 8, 429.
\newblock \url{https://doi.org/10.1063/1.1750679}

\bibitem[{Pavlov {et~al.}(2000)Pavlov, Kasting, Brown, Rages, \&
  Freedman}]{pavlov2000}
Pavlov, A.~A., Kasting, J.~F., Brown, L.~L., Rages, K.~A., \& Freedman, R.
  2000, Journal of Geophysical Research: Planets, 105, 11981.
\newblock
  \url{https://agupubs.onlinelibrary.wiley.com/doi/abs/10.1029/1999JE001134}

\bibitem[{Perrin \& Hartmann(1989)}]{PERRIN1989311}
Perrin, M., \& Hartmann, J. 1989, Journal of Quantitative Spectroscopy and
  Radiative Transfer, 42, 311 .
\newblock
  \url{http://www.sciencedirect.com/science/article/pii/0022407389900770}

\bibitem[{Pierrehumbert(2010)}]{pierrehumbert2010}
Pierrehumbert, R.~T. 2010, Principles of planetary climate (Cambridge
  University Press)

\bibitem[{Pollack {et~al.}(1993)Pollack, Dalton, Grinspoon, Wattson, Freedman,
  Crisp, Allen, Bezard, DeBergh, Giver, {et~al.}}]{pollack1993}
Pollack, J.~B., Dalton, J.~B., Grinspoon, D., {et~al.} 1993, Icarus, 103, 1

\bibitem[{{Press} {et~al.}(1992){Press}, Teukolsky, Vetterling, \&
  Flannery}]{nr}
{Press}, W.~H., Teukolsky, S.~A., Vetterling, W.~T., \& Flannery, B.~P. 1992,
  {Numerical Recipes in Fortran 77}, 2nd edn.

\bibitem[{Ramirez {et~al.}(2014)Ramirez, Kopparapu, Zugger, Robinson, Freedman,
  \& Kasting}]{ramirez2014}
Ramirez, R.~M., Kopparapu, R., Zugger, M.~E., {et~al.} 2014, Nature Geoscience,
  7, 59.
\newblock \url{https://doi.org/10.1038/ngeo2000}

\bibitem[{{Rauer} {et~al.}(2011){Rauer}, {Gebauer}, {Paris}, {Cabrera},
  {Godolt}, {Grenfell}, {Belu}, {Selsis}, {Hedelt}, \& {Schreier}}]{rauer2011}
{Rauer}, H., {Gebauer}, S., {Paris}, P.~V., {et~al.} 2011, \aap, 529, A8

\bibitem[{Robinson \& Crisp(2018)}]{ROBINSON2018}
Robinson, T.~D., \& Crisp, D. 2018, Journal of Quantitative Spectroscopy and
  Radiative Transfer, 211, 78 .
\newblock
  \url{http://www.sciencedirect.com/science/article/pii/S0022407317305101}

\bibitem[{{Rogelj} {et~al.}(2012){Rogelj}, {Meinshausen}, \&
  {Knutti}}]{rogelj2012}
{Rogelj}, J., {Meinshausen}, M., \& {Knutti}, R. 2012, Nature Climate Change,
  2, 248

\bibitem[{Rothman {et~al.}(2010)Rothman, Gordon, Barber, Dothe, Gamache,
  Goldman, Perevalov, Tashkun, \& Tennyson}]{hitemp}
Rothman, L., Gordon, I., Barber, R., {et~al.} 2010, Journal of Quantitative
  Spectroscopy and Radiative Transfer, 111, 2139 , xVIth Symposium on High
  Resolution Molecular Spectroscopy (HighRus-2009).
\newblock
  \url{http://www.sciencedirect.com/science/article/pii/S002240731000169X}

\bibitem[{Scheucher {et~al.}(2018)Scheucher, Grenfell, Wunderlich, Godolt,
  Schreier, \& Rauer}]{Scheucher_2018}
Scheucher, M., Grenfell, J.~L., Wunderlich, F., {et~al.} 2018, The
  Astrophysical Journal, 863, 6.
\newblock \url{https://doi.org/10.3847%2F1538-4357%2Faacf03}

\bibitem[{Schreier(2006)}]{Schreier2006}
Schreier, F. 2006, 174, 783

\bibitem[{{Schreier} {et~al.}(2014){Schreier}, {Gimeno Garc{\'{\i}}a},
  {Hedelt}, {Hess}, {Mendrok}, {Vasquez}, \& {Xu}}]{schreier2014}
{Schreier}, F., {Gimeno Garc{\'{\i}}a}, S., {Hedelt}, P., {et~al.} 2014, JQSRT,
  137, 29

\bibitem[{Schreier {et~al.}(2018{\natexlab{a}})Schreier, Milz, Buehler, \& von
  Clarmann}]{schreier2018agk}
Schreier, F., Milz, M., Buehler, S.~A., \& von Clarmann, T. 2018{\natexlab{a}},
  211, 64

\bibitem[{Schreier {et~al.}(2018{\natexlab{b}})Schreier, St\"adt, Hedelt, \&
  Godolt}]{schreier2018ace}
Schreier, F., St\"adt, S., Hedelt, P., \& Godolt, M. 2018{\natexlab{b}},
  Molec.\ Astrophysics, 11, 1

\bibitem[{{Segura} {et~al.}(2003){Segura}, {Krelove}, {Kasting}, {Sommerlatt},
  {Meadows}, {Crisp}, {Cohen}, \& {Mlawer}}]{segura2003}
{Segura}, A., {Krelove}, K., {Kasting}, J.~F., {et~al.} 2003, Astrobiology, 3,
  689

\bibitem[{{Segura} {et~al.}(2010){Segura}, {Walkowicz}, {Meadows}, {Kasting},
  \& {Hawley}}]{segura2010}
{Segura}, A., {Walkowicz}, L.~M., {Meadows}, V., {Kasting}, J., \& {Hawley}, S.
  2010, Astrobiology, 10, 751

\bibitem[{Seiff {et~al.}(1985)Seiff, Schofield, Kliore, Taylor, Limaye,
  Revercomb, Sromovsky, Kerzhanovich, Moroz, \& Marov}]{seiff1985}
Seiff, A., Schofield, J., Kliore, A., {et~al.} 1985, Advances in Space
  Research, 5, 3 .
\newblock
  \url{http://www.sciencedirect.com/science/article/pii/0273117785901978}

\bibitem[{{Shardanand} \& {Rao}(1977)}]{shardanand1977}
{Shardanand}, \& {Rao}, A.~D.~P. 1977, {Absolute Rayleigh scattering cross
  sections of gases and freons of stratospheric interest in the visible and
  ultraviolet regions}

\bibitem[{Sharp \& Burrows(2007)}]{sharp2007}
Sharp, C.~M., \& Burrows, A. 2007, The Astrophysical Journal Supplement Series,
  168, 140.
\newblock \url{https://doi.org/10.1086%2F508708}

\bibitem[{Sing {et~al.}(2016)Sing, Fortney, Nikolov, Wakeford, Kataria, Evans,
  Aigrain, Ballester, Burrows, Deming, {et~al.}}]{sing2016}
Sing, D.~K., Fortney, J.~J., Nikolov, N., {et~al.} 2016, Nature, 529, 59

\bibitem[{Sneep \& Ubachs(2005)}]{sneep2005}
Sneep, M., \& Ubachs, W. 2005, Journal of Quantitative Spectroscopy and
  Radiative Transfer, 92, 293 .
\newblock
  \url{http://www.sciencedirect.com/science/article/pii/S0022407304002973}

\bibitem[{Spiegel \& Burrows(2010)}]{spiegel2010}
Spiegel, D.~S., \& Burrows, A. 2010, The Astrophysical Journal, 722, 871.
\newblock \url{https://doi.org/10.1088%2F0004-637x%2F722%2F1%2F871}

\bibitem[{Stamnes {et~al.}(1988)Stamnes, Tsay, Wiscombe, \&
  Jayaweera}]{stamnes1988}
Stamnes, K., Tsay, S.-C., Wiscombe, W., \& Jayaweera, K. 1988, Applied optics,
  27, 2502

\bibitem[{{Tabataba-Vakili} {et~al.}(2016){Tabataba-Vakili}, {Grenfell},
  {Grie{\ss}meier}, \& {Rauer}}]{tabataba2016}
{Tabataba-Vakili}, F., {Grenfell}, J.~L., {Grie{\ss}meier}, J.-M., \& {Rauer},
  H. 2016, \aap, 585, A96

\bibitem[{Takagi {et~al.}(2010)Takagi, Suzuki, Sagawa, Baron, Mendrok, Kasai,
  \& Matsuda}]{Takagi2010}
Takagi, M., Suzuki, K., Sagawa, H., {et~al.} 2010, 115, E06014

\bibitem[{Tennyson {et~al.}(2016)Tennyson, Yurchenko, Al-Refaie, Barton, Chubb,
  Coles, Diamantopoulou, Gorman, Hill, Lam, Lodi, McKemmish, Na, Owens,
  Polyansky, Rivlin, Sousa-Silva, Underwood, Yachmenev, \& Zak}]{exomol}
Tennyson, J., Yurchenko, S.~N., Al-Refaie, A.~F., {et~al.} 2016, Journal of
  Molecular Spectroscopy, 327, 73 , new Visions of Spectroscopic Databases,
  Volume II.
\newblock
  \url{http://www.sciencedirect.com/science/article/pii/S0022285216300807}

\bibitem[{Tonkov {et~al.}(1996)Tonkov, Filippov, Bertsev, Bouanich, Van-Thanh,
  Brodbeck, Hartmann, Boulet, Thibault, \& Le~Doucen}]{tonkov1996}
Tonkov, M., Filippov, N., Bertsev, V., {et~al.} 1996, Applied optics, 35, 4863

\bibitem[{Toon {et~al.}(1989)Toon, McKay, Ackerman, \& Santhanam}]{toon1989}
Toon, O.~B., McKay, C.~P., Ackerman, T.~P., \& Santhanam, K. 1989, Journal of
  Geophysical Research: Atmospheres, 94, 16287.
\newblock
  \url{https://agupubs.onlinelibrary.wiley.com/doi/abs/10.1029/JD094iD13p16287}

\bibitem[{Tsang {et~al.}(2008)Tsang, Irwin, Wilson, Taylor, Lee, de~Kok,
  Drossart, Piccioni, Bezard, \& Calcutt}]{tsang2008}
Tsang, C. C.~C., Irwin, P. G.~J., Wilson, C.~F., {et~al.} 2008, Journal of
  Geophysical Research: Planets, 113, doi:10.1029/2008JE003089.
\newblock
  \url{https://agupubs.onlinelibrary.wiley.com/doi/abs/10.1029/2008JE003089}

\bibitem[{Tsiaras {et~al.}(2019)Tsiaras, Waldmann, Tinetti, Tennyson, \&
  Yurchenko}]{Tsiaras2019}
Tsiaras, A., Waldmann, I.~P., Tinetti, G., Tennyson, J., \& Yurchenko, S.~N.
  2019, Nature Astronomy, doi:10.1038/s41550-019-0878-9.
\newblock \url{https://doi.org/10.1038/s41550-019-0878-9}

\bibitem[{Turbet {et~al.}(2019)Turbet, Bolmont, Ehrenreich, Gratier, Leconte,
  Selsis, Hara, \& Lovis}]{turbet2019}
Turbet, M., Bolmont, E., Ehrenreich, D., {et~al.} 2019, arXiv preprint
  arXiv:1911.08878

\bibitem[{Vardavas \& Carver(1984)}]{VARDAVAS1984}
Vardavas, I., \& Carver, J. 1984, Planetary and Space Science, 32, 1307 .
\newblock
  \url{http://www.sciencedirect.com/science/article/pii/0032063384900746}

\bibitem[{von Clarmann {et~al.}(2009)von Clarmann, H\"opfner, Kellmann, Linden,
  Chauhan, Funke, Grabowski, Glatthor, Kiefer, Schieferdecker, Stiller, \&
  Versick}]{clarmann2009}
von Clarmann, T., H\"opfner, M., Kellmann, S., {et~al.} 2009, Atmospheric
  Measurement Techniques, 2, 159.
\newblock \url{https://www.atmos-meas-tech.net/2/159/2009/}

\bibitem[{von Paris {et~al.}(2008)von Paris, Rauer, Grenfell, Patzer, Hedelt,
  Stracke, Trautmann, \& Schreier}]{vonParis2008}
von Paris, P., Rauer, H., Grenfell, J.~L., {et~al.} 2008, Planetary and Space
  Science, 56, 1244 .
\newblock
  \url{http://www.sciencedirect.com/science/article/pii/S0032063308001049}

\bibitem[{von Paris {et~al.}(2015)von Paris, Selsis, Godolt, Grenfell, Rauer,
  \& Stracke}]{vonparis2015}
von Paris, P., Selsis, F., Godolt, M., {et~al.} 2015, Icarus, 257, 406 .
\newblock
  \url{http://www.sciencedirect.com/science/article/pii/S0019103515002389}

\bibitem[{von Paris {et~al.}(2013)von Paris, Selsis, Kitzmann, \&
  Rauer}]{vonParis2013}
von Paris, P., Selsis, F., Kitzmann, D., \& Rauer, H. 2013, Astrobiology, 13,
  899, pMID: 24111995.
\newblock \url{https://doi.org/10.1089/ast.2013.0993}

\bibitem[{von Paris {et~al.}(2010)von Paris, Gebauer, Godolt, Grenfell, Hedelt,
  Kitzmann, Patzer, Rauer, \& Stracke}]{vonParis2010}
von Paris, P., Gebauer, S., Godolt, M., {et~al.} 2010, A\&A, 522, A23.
\newblock \url{https://doi.org/10.1051/0004-6361/201015329}

\bibitem[{Wordsworth {et~al.}(2010{\natexlab{a}})Wordsworth, Forget, \&
  Eymet}]{wordsworth2010}
Wordsworth, R., Forget, F., \& Eymet, V. 2010{\natexlab{a}}, Icarus, 210, 992 .
\newblock
  \url{http://www.sciencedirect.com/science/article/pii/S0019103510002320}

\bibitem[{Wordsworth {et~al.}(2010{\natexlab{b}})Wordsworth, Forget, Selsis,
  Madeleine, Millour, \& Eymet}]{wordsworth2010gliese}
Wordsworth, R., Forget, F., Selsis, F., {et~al.} 2010{\natexlab{b}}, Astronomy
  \& Astrophysics, 522, A22

\bibitem[{Wordsworth {et~al.}(2017)Wordsworth, Kalugina, Lokshtanov, Vigasin,
  Ehlmann, Head, Sanders, \& Wang}]{wordsworth2017}
Wordsworth, R., Kalugina, Y., Lokshtanov, S., {et~al.} 2017, Geophysical
  Research Letters, 44, 665.
\newblock
  \url{https://agupubs.onlinelibrary.wiley.com/doi/abs/10.1002/2016GL071766}

\bibitem[{Wordsworth \& Pierrehumbert(2013)}]{wordsworth2013}
Wordsworth, R.~D., \& Pierrehumbert, R.~T. 2013, The Astrophysical Journal,
  778, 154.
\newblock \url{https://doi.org/10.1088%2F0004-637x%2F778%2F2%2F154}

\bibitem[{Wunderlich {et~al.}(2020)Wunderlich, Scheucher, Godolt, Grenfell,
  Schneider, Schreier, \& Rauer}]{wunderlich2019b}
Wunderlich, F., Scheucher, M., Godolt, M., {et~al.} 2020, ApJ in Prep.

\bibitem[{Wunderlich {et~al.}(2019)Wunderlich, Godolt, Grenfell, St\"adt,
  Smith, Gebauer, Schreier, Hedelt, \& Rauer}]{wunderlich2019a}
Wunderlich, F., Godolt, M., Grenfell, J.~L., {et~al.} 2019, A\&A, 624, A49.
\newblock \url{https://doi.org/10.1051/0004-6361/201834504}

\bibitem[{Yang {et~al.}(2016)Yang, Leconte, Wolf, Goldblatt, Feldl, Merlis,
  Wang, Koll, Ding, Forget, \& Abbot}]{yang2016}
Yang, J., Leconte, J., Wolf, E.~T., {et~al.} 2016, The Astrophysical Journal,
  826, 222.
\newblock \url{https://doi.org/10.3847%2F0004-637x%2F826%2F2%2F222}

\end{thebibliography}

%
\appendix

\section{Heat capacities}
There are in total nine heat capacities, $c_p$ [J mol$^{-1}$ K$^{-1}$], considered in our model: Ar, CH$_4$, CO, CO$_2$, H$_2$, H$_2$O, He, N$_2$, and O$_2$. These had been implemented by various researchers over several decades, most recently by Dr. Philipp von Paris in 2011 and used in e.g. \citet{godolt2016,godolt2019,gebauer2017,gebauer2018,keles2018,Scheucher_2018,wunderlich2019a}. Two of the $c_p$ are implemented as temperature independent, $c_{p,\rm Ar}$ = 20.78600, and $c_{p,\rm He}$ = 20.78603 \citep{chase1998}\footnote{https://webbook.nist.gov/chemistry/}. Heat capacities for CO and H$_2$ are calculated based on the empirical formula \citep{deming1931}:
\begin{equation}
    c_p(T) = \frac{7R}{2} + R\frac{T_{ref}}{T}^2 \frac{e^{\frac{T_{ref}}{T}}}{(e^{\frac{T_{ref}}{T}}-1)^2},
\end{equation}
with the gas constant $R$, and the reference temperatures $T_{ref,\rm CO}$~=~3090~K and $T_{ref,\rm H_2}$~=~6100~K, respectively. Other heat capacities are calculated using the Shomate equation \citep{parks1940}:
\begin{equation}\label{eq:shomate}
    c_p\left(\frac{T}{T_s}\right) = s_1 + s_2\cdot\left(\frac{T}{T_s}\right) + s_3\cdot\left(\frac{T}{T_s}\right)^2 + s_4\cdot\left(\frac{T}{T_s}\right)^3 + s_5\cdot\left(\frac{T}{T_s}\right)^{-2},
\end{equation}
with parameters $s_i$ and temperatures $T_{s}$ for different molecules shown in Table \ref{tbl:shomate}. For CH$_4$ Eq.~\ref{eq:shomate} is only used above 300K within the respective validity range of parameters in Tbl.~\ref{tbl:shomate}. Below we use the linear interpolation: 
\begin{equation}\label{eq:gurvich}
   c_{p,CH_4}(T) = \frac{c_{p,2}-c_{p,1}}{T_2-T_1}(T-T_1) + c_{p,1} 
\end{equation}
to calculations from \citet{gurvich1989} of heat capacities $c_{p,1}$ and $c_{p,2}$ at temperatures $T_1$ and $T_2$, respectively, shown in Table \ref{tbl:gurvich}.

\begin{deluxetable}{c c c c c c}
    \tabletypesize{\footnotesize}
    \tablecolumns{6}
    \tablecaption{Parameters $s_i$ and $T_s$ for heat capacity calculations with the Shomate Equation (eq. \ref{eq:shomate}) together with the respective validity range. \label{tbl:shomate}}
    \tablehead{\colhead{coef.} & \colhead{CH$_4$\tablenotemark{a}} & \colhead{CO$_2$\tablenotemark{b}} & \colhead{H$_2$O\tablenotemark{a}} & \colhead{N$_2$} & \colhead{O$_2$}}    
    \startdata
        $s_1$ & -0.703 & 7.7 & 30.092 & 6.76 & 8.27 \\
        $s_2$ & 108.477 & 5.3$\cdot$ 10$^{-3}$ & 6.832 & 6.06$\cdot$ 10$^{-4}$ & 2.58$\cdot$ 10$^{-4}$ \\
        $s_3$ & -42.522 & -8.3$\cdot$ 10$^{-7}$ & 6.793 & 1.3$\cdot$ 10$^{-7}$ & -- \\
        $s_4$ & 5.863 & -- & -2.534 & -- & -- \\
        $s_5$ & 0.679 & -- & 0.082 & -- & 1.877$\cdot$ 10$^5$ \\
        $T_s$ [K] & 1000 & 1 & 1000 & 1 & 1 \\
        valid [K] & 298-1300 & & 500-1700 & & \\
    \enddata
    \tablenotetext{a}{\citet{chase1998} \url{https://webbook.nist.gov/chemistry/}}
    \tablenotetext{b}{\citet{kasting1991}}
\end{deluxetable}
\begin{deluxetable}{c c c c c}
    \tabletypesize{\footnotesize}
    \tablecolumns{5}
    \tablecaption{CH$_4$ heat capacities $c_{p,i}$ [J mol$^{-1}$ K$^{-1}$] from \citet{gurvich1989}, used in eq. \ref{eq:gurvich}. \label{tbl:gurvich}}
    \tablehead{\colhead{regime} & \colhead{$T_1$[K]} & \colhead{$T_2$[K]} & \colhead{$c_{p,1}$} & \colhead{$c_{p,2}$}}    
    \startdata
        100-200K & 100 & 200 & 33.28 & 33.51 \\
        200-300K & 200 & 300 & 33.51 & 35.76 \\
    \enddata
\end{deluxetable}
\end{document}